\documentclass[a4paper,fleqn]{cas-sc}
 
\usepackage[numbers]{natbib}

\usepackage{url}
\usepackage{enumerate}
\usepackage{graphicx}
\usepackage{array}
\usepackage{amsmath}
\usepackage{epsfig}
\usepackage{amsxtra}
\usepackage{amsbsy}
\usepackage{amsopn}
\usepackage{amsfonts}
\usepackage{amssymb}
\usepackage{mathrsfs}
\usepackage{epsf}
\usepackage{subfigure}
\usepackage{psfrag}
\usepackage{setspace}
\usepackage{xspace}
\usepackage{multirow}
\usepackage{algorithmic}
\usepackage{epstopdf}
\usepackage{color}

\newtheorem{theorem}{Theorem}[section]

\newtheorem{lemma}{Lemma}[section]

\newtheorem{definition}{Definition}[section]

\def\tsc#1{\csdef{#1}{\textsc{\lowercase{#1}}\xspace}}
\tsc{WGM}
\tsc{QE}
\tsc{EP}
\tsc{PMS}
\tsc{BEC}
\tsc{DE}

\begin{document}
\let\WriteBookmarks\relax
\def\floatpagepagefraction{1}
\def\textpagefraction{.001}
\shorttitle{Cohen-Grossberg neural networks with unpredictable and Poisson stable dynamics}
\shortauthors{M. Akhmet et~al.}

\title [mode = title]{Cohen-Grossberg neural networks with unpredictable and Poisson stable dynamics}

\author[1]{Marat Akhmet}[type=editor,
                        auid=000,bioid=1,
                        orcid=0000-0002-2985-286X]
\cormark[1]

\ead{marat@metu.edu.tr}

\address[1]{Department of Mathematics, Middle East Technical University, 06800, Ankara, Turkey}

\author[2,3]{Madina Tleubergenova}

\address[2]{Department of Mathematics, K. Zhubanov Aktobe Regional University, 030000, Aktobe, Kazakhstan}

\address[3]{Institute of Information and Computational Technologies, 050010, Almaty,  Kazakhstan}

\author%
[1] {Akylbek Zhamanshin}

\cortext[cor1]{Corresponding author}

\begin{abstract}
In this paper, Cohen-Grossberg neural networks with unpredictable and compartmental periodic unpredictable strengths of connectivity between cells and inputs are investigated. To approve Poisson stability and unpredictability in neural networks, the method of included intervals and contraction mapping principle are used. The existence, uniqueness, and exponential stability of unpredictable and Poisson stable outputs are discussed. Examples with numerical simulations that support the theoretical results are provided. The dependence of the neural network dynamics on the numerical characteristic, the degree of periodicity, is shown. 
\end{abstract}


\begin{keywords} Cohen-Grossberg neural networks \sep Unpredictable inputs and outputs\sep Poisson stable inputs and outputs \sep Compartmental periodic unpredictable inputs \sep Exponential stability  \sep Numerical simulations 
\end{keywords}

\maketitle

\section{Introduction}\label{introduction}

Cohen–Grossberg neural networks (CGNNs) were first proposed by Cohen and Grossberg in 1983 \cite{Cohen}. The class of networks has intensive applications within various engineering and scientific fields such as neuro-biology, population biology, and computing technology. Such applications strongly depend on the dynamic behavior of networks, so the analysis of the dynamics of the model is necessary.

As is known, CGNNs include many well-known neural networks, such as the Lotka-Volterra system, cellular neural networks, Hopfield neural networks,  and are described as follows:
\begin{eqnarray} \label{cohen1} 
	&&x'_{i}(t)=-a_{i}(x_{i}(t))\Big[b_i(x_i(t))-\sum_{j=1}^{n} c_{ij}(t)f_j(x_{j}(t)) + v_{i}(t)\Big],
\end{eqnarray}
where $i=1,2,\cdots,n,$ is the number of neurons; $x_i(t)$ is the state of $i$th neuron at time $t;$ $a_i(x_i(t))$ is an amplification function; $b_i(x_i(t))$ is the rate with which the unit self-regulates or resets its potential, when isolated from other units and inputs; $c_{ij}(t)$ is the strengths of connectivity between cell $i$ and $j$ at time $t;$ the function $v_i(t)$ is an external input source introduced from outside the network to cell $i$ at time $t.$

In recent decades, scientists have been investigating the dynamics of modified CGNNs. Of great interest to researchers are oscillations with recurrence. Therefore, periodic, almost-periodic solutions of CGNNs are deeply studied \cite{KongF,Cai,Liang,Meng,Zhao,LiFan}. For instance, in paper \cite{Kong}, using the approximation technique, periodic and homoclinic solutions of Cohen–Grossberg neural networks with time-varying delays were investigated. The following model, where the external inputs are discontinuous periodic is considered,
\begin{eqnarray}
	\frac{dx_i(t)}{dt}=a_i(x_i(t))\Big[-b_i(t)x_i(t)+\sum_{j=1}^{n} c_{ij}(t)g_j(x_{j}(t)) +\sum_{j=1}^{n} d_{ij}(t)g_j(x_{j}(t-\tau_{ij}(t))) + I_{i}(t)\Big], \quad i=1,2,\ldots,n.
\end{eqnarray}

By means of functional differential inclusions, in the article \cite{WangHuang}, the periodicity and multi periodicity of CGNNs with discontinuous right-hand sides and time-varying and infinite delays were studied. The authors assumed that in the following neural network, all coefficients and input data are continuous periodic functions,
\begin{eqnarray}\label{delays}
		\frac{dx_i(t)}{dt}=&q_i(x_i(t))\Big[-d_i(t)x_i(t)+\sum_{j=1}^{n} a_{ij}(t)f_j(x_{j}(t)) +\sum_{j=1}^{n} b_{ij}(t)f_j(x_{j}(t-\tau(t))) \nonumber \\
	&+\sum_{j=1}^{n} c_{ij}(t)\int_{0}^{\infty}f_j(x_{j}(t-s))l_j(s)ds+ I_{i}(t)\Big], \quad i=1,2,\ldots,n.
\end{eqnarray}
Several improved criteria  to derive positive solutions with $\omega-$periodicity and $\omega-$multi-periodicity for CGNNs (\ref{delays}) with impulses are given
in \cite{CaiWangHuang}.

In \cite{LiSong}, the authors obtained sufficient conditions to verify the existence, exponential stability and stabilization of periodic solutions of CGNNs with impulses,
\begin{eqnarray*}
	\begin{cases}
		\displaystyle x'_i(t)=-\alpha_i(x_i)\Big[-\beta_i(t,x_i)+\sum_{j=1}^{n} \Big(a_{ij}(t)f_j(x_{j}) + b_{ij}(t)g_j(x_{j})\Big) + J_{i}(t)\Big], &t\geq 0, t\neq t_k,\\
		\displaystyle \Delta x_i(t_k)=x_i(t_k)-x_i(t^-_k)=I_{ik}(t_k,x_i(t^-_k)), &k\in \mathbb{N},i=1,2,\ldots,n.
	\end{cases}
\end{eqnarray*}

The most sophisticated recurrent functions are Poisson stable ones \cite{Poin,AkhmetMU}. But we have not found Poisson stable oscillations of CGNNs in the literature. Perhaps because previously known methods of confirming Poisson stability are not so easy for comprehension and applications. To simplify this task, in papers  \cite{Miskolc,Compart}, we have proposed a relation between intervals of convergence for inputs and outputs of models. It is called \textit{the method of included intervals}, and has been successfully applied to the study of Poisson stable motions in neural networks \cite{AkhmetSymmetry,Inert2}.

In order to strengthen the role of recurrence as a chaotic ingredient, the Poisson stability was extended to the unpredictability property \cite{Akhmet2016}. Thus, the concept of unpredictable functions was introduced. Moreover, any system which admits unpredictable solution has Poincare chaos \cite{Akhmet2016}. In papers \cite{Kagan1,Kagan2,Kagan3}, the synchronization of Poincare chaos in semiconductor gas discharge models was considered. Currently, Poisson stable and unpredictable oscillations of Hopfield type neural networks \cite{AkhmetHopfield,Akhmet7}, shunting inhibitory cellular neural networks \cite{AkhmetSymmetry,Akhmet2013}, and inertial neural networks \cite{Inert2}, have been investigated. 

The remainder of this paper is structured as follows. In the next section, the basic definitions are given. The theorem on the unpredictability of a compartmental periodic unpredictable functions is announced. The conditions for neural networks that are sufficient to obtain the results of the article are presented. The neural network (\ref{cohen1}) reduced to a quasi-linear model, which, in turn, is convenient for investigating the existence of a unique solution and its stability. In Sections \ref{results} and \ref{compart1} it is proved that unpredictable and Poisson stable motions take place in the dynamics of the CGNNs (\ref{cohen1}) when the strengths of connectivity between cells and inputs are Poisson stable, unpredictable or compartmental periodic unpredictable. Moreover, it is shown that the solutions are  exponentially stable. Section \ref{numexmple} contains the numerical examples that confirm the feasibility of theoretical results. The section is closed with examples of CGNNs (\ref{cohen1}), where  the strengths of connectivity between cells and inputs are compartmental periodic unpredictable functions. Finally, prospects of the obtained results for chaos control and synchronization in neural networks are discussed in \textit{Conclusions}.

\section{Reduction of the model to a quasilinear mode}\label{reduction}

In this section, in order to use the methods of the first approximation, by an integral transformation we reduce a strongly nonlinear model (\ref{cohen1}) to a quasi-linear system. Throughout the paper, we will use  the norm $\displaystyle \left\|g\right\|=\max_{i=1,2, \ldots,n}\left|g_{i}\right|,$  where $\left|\cdot\right|$ is the absolute value.

Let us commence with the definitions of Poisson stable and unpredictable functions.
\begin{definition}  \label{1def1} \cite{sell} A bounded function $g(t):\mathbb{R}\rightarrow\mathbb{R}^n$   is called Poisson stable if there exists a sequence $t_p,$ $t_p\rightarrow \infty$ as $p \rightarrow \infty,$ such that $\|g(t+t_{p})-g(t)\|\rightarrow 0$ uniformly on compact subsets of $\mathbb{R}.$ 
\end{definition}

\begin{definition}  \label{ps} \cite{Akhmet2016}	A  bounded function $g : \mathbb  R  \rightarrow \mathbb  R^n$ is said to be unpredictable if there exist positive numbers $\epsilon_{0}, \delta$ and sequences $t_{p}\rightarrow\infty,$ $s_{p}\rightarrow\infty$ as $p \rightarrow\infty,$ such that $\|g(t+t_{p})-g(t)\|\rightarrow 0$ uniformly on compact subsets of $\mathbb  R$ and $\|g(t+t_{p})-g(t)\|>\epsilon_{0}$ for each $t\in [s_{p}-\delta, s_{p}+\delta]$ and $ p\in \mathbb  N$. 
\end{definition}

The sequence  $t_p,p=1,2,\ldots,$ in  Definitions \ref{1def1},\ref{ps} is called \textit{ the convergence sequence},  and correspondingly we shall say about   \textit{the convergence property}, while the existence of  positive numbers $\epsilon_{0}, \delta$ and  sequence $s_p$  is said to be \textit{the separation property}.

It is easily seen, reading the last two definitions, that all  unpredictable functions  make a subset of  Poisson stable functions  specified with an  additional property of  separation. It was proved in \cite{Akhmet2016} that the property  guarantees  chaotic dynamics of the unpredictable motion. Loosely speaking, one can say that an  unpredictable function is a  Poisson stable function with assigned chaotic behavior.

In paper \cite{Compart}, unpredictable functions that combine periodic dynamics with the unpredictability were considered. These functions are called \textit{compartmental periodic unpredictable}, and allow to see new types of oscillations. The definition of compartmental periodic unpredictable function is as follows:
\begin{definition} \cite{Compart} \label{compart} A   function $f(t): \mathbb  R  \rightarrow \mathbb  R^n$ is said to be a \textit{compartmental periodic unpredictable} function if $f(t) = G(t,t),$  where $G(u,v)$  is a   continuous bounded function, periodic in  $u$ uniformly with respect to $v,$ and unpredictable in $v$ uniformly with respect to $u,$ i.e., there exist positive numbers $\omega,$ $\epsilon_0$, $\delta$ and sequences $t_p$, $s_p$, both of which diverge to infinity, such that $\displaystyle G(u+\omega,v)=G(u,v)$ for all $u,v \in \mathbb{R},$ $\displaystyle \sup_{u \in \mathbb{R}} \|G(u,v+t_p)-G(u,v)\| \rightarrow 0$  as $p\rightarrow\infty$ uniformly on bounded intervals of $v,$  and  $\displaystyle \|G(u,v+t_p)-G(u,v)\|> \epsilon_0$ for  $v\in [s_p-\delta, s_p+\delta],$ $u\in \mathbb{R}$ and $p\in\mathbb{N}$.
\end{definition}


Let us consider convergence sequence $t_p,$  and fix a positive number $\omega.$ We shall call a  number $\tau_{\omega}$  as \textit{Poisson shift} of the sequence $t_p$ with respect to  $\omega$  if  there exist a subsequence $t_{p_l}$ of the sequence such that $t_{p_l} \rightarrow \tau_{\omega} (mod \ \omega)$ as $l \rightarrow \infty.$ Denote by $\mathcal{T}_{\omega}$  the set of all Poisson shifts. The set $\mathcal{T}_{\omega}$ is not empty, it can consist of several or even an infinite number of elements. The~number $\kappa_{\omega}=inf \ \mathcal{T}_{\omega},$  $0\leq \kappa_{\omega}<\omega,$ is called \textit{Poisson number with respect to the number $\omega.$} If $\kappa_{\omega}=0,$ then we say that the sequence $t_p$ satisfies \textit{kappa property with respect to the number $\omega.$}

The unpredictability of compartmental functions is confirmed by the following theorem.

\begin{theorem} \cite{Compart} \label{lemma333} Assume that a continuous and bounded function $G(u,v):\mathbb{R}\times \mathbb{R} \rightarrow \mathbb{R}^n,$ is $\omega-$periodic in $u.$ The function  $f(t)=G(t,t)$ is unpredictable if the following conditions are~valid,
	\begin{enumerate}
		\item[(i)] for each $\epsilon>0$ there exists a positive number $\eta$ such that $\|G(t+s,t)-G(t,t)\|<\epsilon$ if $|s|<\eta, t \in \mathbb{R};$
	\end{enumerate}
	there exist sequences $t_p,$ $s_p$ both of which diverges to infinity as $p\rightarrow \infty,$ and positive numbers $\epsilon_{0}, \delta$, such~that
	\begin{enumerate}
		\item[(ii)]	the sequence $t_p$ satisfies kappa property with respect to the period $\omega;$	
		\item[(iii)] $\|G(t,t+t_{p})-G(t,t)\|\rightarrow 0,$ uniformly on each bounded interval  $I$ of $t;$ 
		\item[(iv)] $\displaystyle \inf_{[s_p-\delta, s_p+\delta]} \|G(t,t+t_{p})-G(t,t)\|>\epsilon_0,$ $p \in \mathbb{N}.$
	\end{enumerate}
\end{theorem}

It is obvious that conditions $(i)-(iv)$ of the last theorem are weaker than those in Definition  \ref{ps}, but they are easily verifiable.

In this paper, we consider CGNNs (\ref{cohen1}), provided that a solution $x(t)$ is bounded such that $\sup_{t \in \mathbb{R}} \|x(t)\|< H_0,$  where $H_0$ is a fixed positive number.

The following assumptions are needed throughout this paper,
\begin{itemize}
	\item[(C1)]	each $a_i(s),$ $i=1,2,\ldots,n,$ $|s|<H_0,$ is continuous and there exist positive numbers $\underline{a}_i$ and $\bar{a_i}$  such that $\underline{a}_i\le a_i(s) \le \bar{a_i},$  $i=1,2,\ldots,n;$
	\item[(C2)] functions $b_i(s),$ $i=1,2,\ldots,n,$ are continuous;
	\item[(C3)] functions $f_i,$ $i=1,2,\ldots,n,$ are Lipschitzian with constants $L^f_i,$ $|f_i(s_1)-f_i(s_2)|\le L^f_i|s_1-s_2|$ for all $|s_1|<H_0,$ $|s_2|<H_0;$
	\item[(C4)] inputs $v_i(t)$ and strengths of connectivity between cells $c_{ij}(t),$ $i=1,2,\ldots,n,$ $j=1,2,\ldots,n,$ are unpredictable with common sequences of convergence $t_p$  and separation sequence  $s_p,$ $p=1,2,\ldots.$
\end{itemize}

Condition (C1) implies that  for each $i=1,2,\ldots,n,$ there exists a function $h_i(s)$ such that $h_i(s)=\int_{0}^{s}\frac{1}{a_i(\tau)}d \tau,$ $h_i(0)=0.$ Obviously, $h'_i(s)=1/a_i(s).$ By $a_i(s)>0,$ we obtain that $h_i(s)$ is increasing in $s$ and the inverse function $(h_i)^{-1}(s)$  is existential, continuous, and differential. So, $(h_i^{-1})'(s)=a_i(s),$ where $(h_i^{-1})'$ is the derivative of function $(h_i^{-1})(s)$ in $s.$ 

It is not difficult to see that 
\begin{eqnarray*}		
	&&|h_i^{-1}(s_1)-h_i^{-1}(s_2)|=(h_i^{-1})'(\zeta)(s_1-s_2)|=|a_i(\zeta)||s_1-s_2|, \quad i=1,2,\cdots,n,
\end{eqnarray*}
for $|s_1|<H_0,$ $|s_2|<H_0,$ and $\zeta$ between $s_1,$ $s_2.$ This is why, the function $h_i^{-1}(s)$ satisfies Lipschitz conditions
\begin{eqnarray}\label{h2}		
	&&\underline{a}_i|s_1-s_2|\le|h_i^{-1}(s_1)-h_i^{-1}(s_2)|\le \bar{a_i}|s_1-s_2|, \quad i=1,2,\ldots,n,
\end{eqnarray}
if $|s_1|<H_0,$ $|s_2|<H_0.$

We propose the following transformation, which reduces the highly non-linear model to a quasilinear one,
\begin{eqnarray}\label{subst}
	y_i(t)=h_i(x_i(t)), \quad i=1,2,\ldots,n.
\end{eqnarray}
One  has that $\displaystyle y'_i(t)=h'_i(x_i(t))x'_i(t)=\frac{x'_i(t)}{a_i(x_i(t))}$ and $x_i(t)=h^{-1}_i(y_i(t)).$ Since $\|x(t)\|<H_0,$ then  $\|y(t)\|< H,$ where $\displaystyle H=\frac{H_0}{\max_{i}\bar{a_i}},$ for all $t \in \mathbb{R}.$  

Using substitution (\ref{subst}) for the system (\ref{cohen1}), we get that 
%

\begin{eqnarray} \label{cohen3} 
	&&y'_{i}(t)=-u_{i}(y_i(t))y_{i}(t)+\sum_{j=1}^{n} c_{ij}(t)f_j(h_j^{-1}(y_{j}(t))) + v_{i}(t),
\end{eqnarray}
where $\displaystyle u_i(s)=\frac{b_i(h^{-1}_i(s))}{s},$ $i=1,2,\ldots,n.$

For the sake of simplicity, we will use notations:
\begin{eqnarray*}
	\displaystyle m^{c}_{ij}= \sup_{t \in \mathbb{R}}|c_{ij}(t)| , \quad 	m^{f}_{i}=\sup_{t \in \mathbb{R}}| f_{i}(t)|, \quad  m^{v}_{i}=\sup_{t \in \mathbb{R}}| v_{i}(t)|,
\end{eqnarray*}
for each $i=1,2,\ldots,n; j=1,2,\ldots,n.$

Assume that  for all $i=1,2,\ldots,n,$ the following conditions are satisfied,
\begin{itemize}
	\item[(C5)]	there exist positive numbers $m_i, M_i$ such that $0<m_i\le u_i(s)\le M_i,$ for  $|s|<H;$ 
	\item[(C6)] there exists positive number $L^u_i$ such that $|u_i(s_1)-u_i(s_2)|\le L^u_i|s_1-s_2|$ for all $|s_1|<H,$ $|s_2|<H;$  
	\item[(C7)] $\displaystyle \frac{1}{m_i}\Big(\sum_{j=1}^{n}m^c_{ij}m^f_i+m^v_{i}\Big)<H;$   
	\item[(C8)] $\displaystyle \frac{1}{m_i}\Big(\frac{L^u_i}{m_i}(\sum_{j=1}^{n}m^c_{ij}m^f_i+m^v_i)+\sum_{j=1}^{n}m^c_{ij}L^f_i\bar{a_i}\Big)<1.$  
\end{itemize}

According to the theory of differential equations \cite{Hartman}, the bounded function $y(t)=(y_1(t),y_2(t),\cdots,y_n(t)),$ is a solution of system (\ref{cohen3}) if and only if the following equalities
\begin{eqnarray}\label{integral}
	&&	 \displaystyle y_{i}(t)=  \int_{-\infty}^{t} e^{-\int_{s}^{t}u_{i}(y_i(\tau))d\tau}\Big(\displaystyle\sum_{j=1}^{n}c_{ij}(s)f_j(h^{-1}_{j}(y_j(s)))+v_{i}(s)\Big)ds, 
\end{eqnarray} 
are valid for all $i=1,2,\ldots,n.$

\section{Unpredictable and Poisson stable motions}\label{results}

In this part of the paper, the existence of Poisson stable and unpredictable dynamics of the neural network (\ref{cohen1}) is considered. The neural network (\ref{cohen1}) with unpredictable and compartmental periodic unpredictable strengths of connectivity between cells and external inputs is investigated. Using the method of included intervals \cite{Compart} and contraction mapping principle, it is proved that Theorem \ref{theorem1} on the existence exponentially stable Poisson stable solution is valid. Moreover, it is shown by Theorem \ref{theorem2} that the unpredictable motions, which are exponentially stable, are present in the dynamics of the neural network (\ref{cohen1}). 

Denote by $\cal B$ the set of functions $\phi=(\phi_1,\phi_2,\cdots,\phi_n),$  where each $\phi_i,$ $i=1,2,\ldots,n,$ is Poisson stable with convergence sequence $t_p,$ $p=1,2,\cdots,$ and $|\phi_i(t)|<H,$ $t\in \mathbb{R},$ $i=1,2,\cdots,n.$ 

Define on $\cal B$ the operator $T$ such that $T\phi(t)=(T_1\phi(t),T_2\phi(t),\cdots,T_n\phi(t)),$ where

\begin{eqnarray}\label{operator}
	\displaystyle T_i \phi(t)= \int_{-\infty}^{t} e^{-\int_{s}^{t}u_{i}(\phi_i(\tau))d\tau}\Big(\displaystyle\sum_{j=1}^{n}c_{ij}(s)f_j(h^{-1}_{j}(\phi_j(s)))+v_{i}(s)\Big)ds, \quad i=1,2,\ldots,n. 
\end{eqnarray}

\begin{lemma}\label{lemma1} The operator $T$ is invariant in ${\cal B}$ provided that conditions $(C1)-(C7)$ are satisfied.
\end{lemma}

\noindent {\bf Proof.}  Fix a function $\phi \in \cal B.$ We have that
\begin{eqnarray*}\label{operator2}
	&&	 \displaystyle |T_i \phi(t)| \le \int_{-\infty}^{t} e^{-\int_{s}^{t}u_{i}(\phi_i(\tau))d\tau}\Big(\displaystyle\sum_{j=1}^{n}|c_{ij}(s)||f_j(h^{-1}_{j}(\phi_j(s)))|+|v_{i}(s)|\Big)ds \le \\
	&& \int_{-\infty}^{t} e^{-m_i(t-s)}\Big(\displaystyle\sum_{j=1}^{n}m^c_{ij}m^f_i+m^v_{i}\Big)ds \le \frac{1}{m_i}\Big(\displaystyle\sum_{j=1}^{n}m^c_{ij}m^f_i+m^v_{i}\Big),
\end{eqnarray*}
for each $i=1,2,\ldots,n.$ Therefore, condition (C7) implies that $\|T\phi(t)\|_0<H.$

Next, applying the method of included intervals \cite{Compart}, we will show that the sequence of images $T \phi(t+t_p)$ uniformly converges to $T\phi(t)$ as $p \rightarrow \infty$ on compact subsets of $\mathbb{R}.$

Let us fix an arbitrary $\epsilon>0$  and a section $[\alpha,\beta],  -\infty <\alpha<\beta<\infty.$ There exist numbers $\gamma, \xi$ such that $\gamma<\alpha$ and $\xi>0,$ which satisfy the following inequalities:
\begin{equation}\label{eps1} 
	\frac{1}{m_{i}}e^{-m_{i}(\alpha-\gamma)} \Big(\sum_{j=1}^{n} m^c_{ij}(m^f_i+\frac{1}{4}L^f_i\bar{a_i}H)+m^{v}_i\Big)<\frac{\epsilon}{8}, 
\end{equation}
\begin{equation}\label{eps3}
	\frac{L^u_i}{m^2_{i}}\xi\Big(\sum_{j=1}^{n} m^c_{ij} m^f_i+m^{v}_i  \Big)< \frac{\epsilon}{4},
\end{equation}
and \begin{equation}\label{eps4}
	\frac{1}{m_{i}}\xi\Big(\sum_{j=1}^{n} m^c_{ij}L^f_i\bar{a_i}+nm^f_i+1\Big)< \frac{\epsilon}{4},
\end{equation}
for all $i=1,2,\ldots,n.$

Since the functions $v_{i}(t),$ $c_{ij}(t),$  $i,j=1,2,\ldots,p,$ are unpredictable, $\phi(t)$  belongs to $\cal{B},$  and the convergence sequence, $t_p,$ is common to all of them, then the following inequalities are true: $|v_{i}(t+t_p)-v_{i}(t)|<\xi,$ $|c_{ij}(t+t_p)-c_{ij}(t)|<\xi,$   $|\phi_{i}(t+t_p)-\phi_{i}(t)|< \xi$ for $t \in [\gamma,\beta].$ 
We obtain that

\begin{eqnarray*}
	&& |T_{i} \phi(t+t_p)-T_{i} \phi(t)| \le \Big| \int_{-\infty}^t e^{-\int_{s}^{t}u_{i}(\phi_i(\tau+t_p))d\tau}  (\sum_{j=1}^{n} c_{ij}(s+t_p)f_{j}(h^{-1}_{j}(\phi_{j}(s+t_p)))+v_{i}(s+t_p)) ds -\\
	&&  \int_{-\infty}^t e^{-\int_{s}^{t}u_{i}(\phi_i(\tau))d\tau}  (\sum_{j=1}^{n} c_{ij}(s)f_{j}(h^{-1}_{j}(\phi_{j}(s)))+v_{i}(s)) ds\Big| \le  \int_{-\infty}^t \Big|e^{-\int_{s}^{t}u_{i}(\phi_i(\tau+t_p))d\tau}-e^{-\int_{s}^{t}u_{i}(\phi_i(\tau)))d\tau}\Big| \times \\
	&& \Big |\sum_{j=1}^{n} c_{ij}(s+t_p)f_{j}(h^{-1}_{j}(\phi_{j}(s+t_p)))+v_{i}(s+t_p)\Big|ds +\\
	&&\int_{-\infty}^t e^{-\int_{s}^{t}u_{i}(\phi_i(\tau))d\tau}\Big | \sum_{j=1}^{p}\ c_{ij}(s+t_p)\Big(f_{j}(h^{-1}_{j}(\phi_{j}(s+t_p)))-f_{j}(h^{-1}_{j}(\phi_{j}(s)))\Big)+ \\
	&& \sum_{j=1}^{p}\ (c_{ij}(s+t_p)-c_{ij}(s))f_{j}(h^{-1}_{j}(\phi_{j}(s)))+v_{i}(s+t_p)-v_{i}(s)\Big| ds,
\end{eqnarray*}
for each $i=1,2,\ldots,n.$	

Consider the sum of integrals in the last inequality on two intervals, $(-\infty, \gamma]$ and $(\gamma, t].$ Using inequalities (\ref{eps1})--(\ref{eps4}), we get that the following estimates are correct for each $i=1,2,\dots,n:$

\begin{eqnarray*}
	&& I_1=\displaystyle \int_{-\infty}^{\gamma} \Big|e^{-\int_{s}^{t}u_{i}(\phi_i(\tau+t_p))d\tau}-e^{-\int_{s}^{t}u_{i}(\phi_i(\tau)))d\tau}\Big|\Big |\sum_{j=1}^{n} c_{ij}(s+t_p)f_{j}(h^{-1}_{j}(\phi_{j}(s+t_p)))+v_{i}(s+t_p)\Big|ds +\\
	&& \int_{-\infty}^{\gamma} e^{-\int_{s}^{t}u_{i}(\phi_i(\tau))d\tau}\Big | \sum_{j=1}^{p}\ c_{ij}(s+t_p)\Big(f_{j}(h^{-1}_{j}(\phi_{j}(s+t_p)))-f_{j}(h^{-1}_{j}(\phi_{j}(s)))\Big)+ \\
	&& \sum_{j=1}^{p}\ (c_{ij}(s+t_p)-c_{ij}(s))f_{j}(h^{-1}_{j}(\phi_{j}(s)))+v_{i}(s+t_p)-v_{i}(s)\Big| ds \leq \\
	&& \int_{-\infty}^{\gamma}  2e^{-m_{i}(t-s)} \Big(\sum_{j=1}^{n} m^c_{ij}m^f_i+m^{v}_i \Big)ds+  \int_{-\infty}^{\gamma} e^{-m_{i}(t-s)}  \Big(\sum_{j=1}^{n} m^c_{ij}L^f_i\bar{a_i}H+2\sum_{j=1}^{n} m^c_{ij}m^f_i+2m^{v}_i \Big)ds \leq \nonumber \\
	&&\frac{2}{m_{i}}e^{-m_{i}(\alpha-\gamma)} \Big(\sum_{j=1}^{n} m^c_{ij} m^f_i+m^{v}_i  \Big)+\frac{1}{m_{i}}e^{-m_{i}(\alpha-\gamma)}\Big(\sum_{j=1}^{n} m^c_{ij}L^f_i\bar{a_i}H+2\sum_{j=1}^{n} m^c_{ij}m^f_i+2m^{v}_i \Big)ds \leq  \nonumber \\
	&&\frac{4}{m_{i}}e^{-m_{i}(\alpha-\gamma)} \Big(\sum_{j=1}^{n} m^c_{ij}(m^f_i+\frac{1}{4}L^f_i\bar{a_i}H)+m^{v}_i\Big)<\frac{\epsilon}{2},
\end{eqnarray*}
and
\begin{eqnarray*}
	&& I_2=\displaystyle \int_{\gamma}^{t} | e^{-\int_{s}^{t}u_{i}(\phi_i(\tau+t_p))d\tau}-e^{-\int_{s}^{t}u_{i}(\phi_i(\tau))d\tau}|\Big |\sum_{j=1}^{n} c_{ij}(s+t_p)f_{j}(h^{-1}_{j}(\phi_{j}(s+t_p)))+v_{i}(s+t_p)\Big|ds +\\
	&& \int_{\gamma}^{t} e^{-\int_{s}^{t}u_{i}(\phi_i(\tau))d\tau} \Big| \sum_{j=1}^{n}\ c_{ij}(s+t_p)(f_{j}(h^{-1}_{j}(\phi_{j}(s+t_p)))-f_{j}(h^{-1}_{j}(\phi_{j}(s))))+ \\
	&& \sum_{j=1}^{n}\ (c_{ij}(s+t_p)-c_{ij}(s))f_{j}(h^{-1}_{j}(\phi_{j}(s)))+v_{i}(s+t_p)-v_{i}(s)\Big| ds \le \\
	&& \int_{\gamma}^{t} e^{-\int_{s}^{t}u_{i}(\phi_i(\tau))d\tau}\Big|e^{-\int_{s}^{t}(u_{i}(\phi_i(\tau+t_p))-u_{i}(\phi_i(\tau)))d\tau}-1\Big| \Big(\sum_{j=1}^{n} |c_{ij}(s+t_p)||f_{j}(h^{-1}_{j}(\phi_{j}(s+t_p)))|+|v_{i}(s+t_p)|\Big)ds+\\
	&& \int_{\gamma}^{t} e^{-\int_{s}^{t}u_{i}(\phi_i(\tau))d\tau} \Big( \sum_{j=1}^{n}\ |c_{ij}(s+t_p)||f_{j}(h^{-1}_{j}(\phi_{j}(s+t_p)))-f_{j}(h^{-1}_{j}(\phi_{j}(s)))|+ \\
	&& \sum_{j=1}^{n}\ |c_{ij}(s+t_p)-c_{ij}(s)||f_{j}(h^{-1}_{j}(\phi_{j}(s)))|+|v_{i}(s+t_p)-v_{i}(s)|\Big) ds < \\
	&&\int_{\gamma}^{t} e^{-m_i(t-s)}L^u_i(t-s)\sup_{t\in \mathbb{R}}|\phi_i(s+t_p)-\psi_i(s)|\Big(\sum_{j=1}^{n} m^c_{ij}m_f+m^{v}_i \Big)ds+ \nonumber \\
	&& \int_{\gamma}^{t} e^{-m_{i}(t-s)}  \Big(\sum_{j=1}^{n} m^c_{ij}L^f_i\bar{a_i}\xi+n\xi m^f_i+\xi\Big)ds \leq \nonumber \\
	&&\frac{1}{m^2_{i}}L^u_i\xi\Big(\sum_{j=1}^{n} m^c_{ij} m^f_i+m^{v}_i  \Big)+\frac{1}{m_{i}}\xi\Big(\sum_{j=1}^{n} m^c_{ij}L^f_i\bar{a_i}+nm^f_i+1\Big)<\frac{\epsilon}{4}+\frac{\epsilon}{4}=\frac{\epsilon}{2}.
\end{eqnarray*}
This is why, for all $t\in [\alpha,\beta]$ and $i=1,2,\ldots,n,$ we have that $|T_{i}\phi(t+t_{p})-T_{i}\phi(t)|\le I_1+I_2<\epsilon.$ So, the function $T\phi(t+t_p)$ uniformly convergences to $T\phi(t)$ on compact subsets of $\mathbb{R},$ and it is true that $T:\cal B \rightarrow \cal B.$ $\Box$

\begin{lemma} \label{lemma3} Assume that conditions $(C1)-(C8)$ are valid. Then the operator $T$ is contractive in ${\cal B}.$  
\end{lemma}

\noindent {\bf Proof.} 	Let  functions $\phi$ and $\psi$ belong to ${\cal B}.$ For fixed $i=1,2,\dots,n,$ we have  that
\begin{eqnarray*}
	&& |T_{i} \phi(t)-T_{i} \psi(t)| \le \Big|\int_{-\infty}^{t} e^{-\int_{s}^{t}u_{i}(\phi_i(\tau))d\tau}\Big(\displaystyle\sum_{j=1}^{n}c_{ij}(s)f_j(h^{-1}_{j}(\phi_j(s)))+v_{i}(s)\Big)ds-\\
	&&\int_{-\infty}^{t} e^{-\int_{s}^{t}u_{i}(\psi_i(\tau))d\tau}\Big(\displaystyle\sum_{j=1}^{n}c_{ij}(s)f_j(h^{-1}_{j}(\psi_j(s)))+v_{i}(s)\Big)ds\Big| \le\\
	&& \Big|\int_{-\infty}^{t} e^{-\int_{s}^{t}u_{i}(\phi_i(\tau))d\tau}(1-e^{-\int_{s}^{t}(u_{i}(\psi_i(\tau))-u_{i}(\phi_i(\tau)))d\tau})\Big(\displaystyle\sum_{j=1}^{n}c_{ij}(s)f_j(h^{-1}_{j}(\phi_j(s)))+v_{i}(s)\Big)ds\Big|+\\
	&&\Big|\int_{-\infty}^{t} e^{-\int_{s}^{t}u_{i}(\psi_i(\tau))d\tau}\Big(\displaystyle\sum_{j=1}^{n}c_{ij}(s)f_j(h^{-1}_{j}(\psi_j(s)))-\displaystyle\sum_{j=1}^{n}c_{ij}(s)f_j(h^{-1}_{j}(\psi_j(s)))\Big)ds\Big|\le\\
	&&\int_{-\infty}^{t}e^{-m_i(t-s)}L^u_i \sup_{t \in \mathbb{R}}|\phi_i-\psi_i|(t-s)\Big(\displaystyle\sum_{j=1}^{n}|c_{ij}(s)||f_j(h^{-1}_{j}(\phi_j(s)))|+|v_{i}(s)|\Big)ds+\\
	&&\int_{-\infty}^{t} e^{-m_i(t-s)}\Big(\displaystyle\sum_{j=1}^{n}|c_{ij}(s)||f_j(h^{-1}_{j}(\psi_j(s)))|-\displaystyle\sum_{j=1}^{n}|c_{ij}(s)||f_j(h^{-1}_{j}(\psi_j(s)))|\Big)ds\le\\
	&& \frac{1}{m^2_i}L^u_i(\sum_{j=1}^{n}m^c_{ij}m^f_i+m^v_i)\sup_{t \in \mathbb{R}}|\phi_i-\psi_i|+\frac{1}{m_i}\sum_{j=1}^{n}m^c_{ij}L^f_i\bar{a_i}\sup_{t \in \mathbb{R}}|\phi_i-\psi_i|\le\\
	&&\frac{1}{m_i}\Big(\frac{L^u_i}{m_i}(\sum_{j=1}^{n}m^c_{ij}m^f_i+m^v_i)+\sum_{j=1}^{n}m^c_{ij}L^f_i\bar{a_i}\Big) \sup_{t \in \mathbb{R}}|\phi_i-\psi_i|.
\end{eqnarray*}

In accordance with condition (C8), the operator $T$ is  contractive in ${\cal B}.$  $\Box$

Denote $\displaystyle L_u=\max_{(i)} L^u_i,$ $\displaystyle m=\min_{(i)} m_i,$ $\displaystyle m_c=\max_{(i)} \sum_{j=1}^{n}m^c_{ij},$ $\displaystyle L_f=\max_{(i)} L^f_i,$ $\displaystyle m_v=\max_{(i)} m^v_i,$ $\displaystyle \bar{a}=\max_{(i)}\bar{a_i},$ and $\sigma$ is a positive number such that $\sigma<m.$

The following condition will be needed to prove the exponential stability of the solution,
\begin{itemize}
	\item[(C9)]	$HL_u\frac{1}{\sigma}+L_u(m_cm_f+m_v)\frac{1}{\sigma}\Big(\frac{2}{m-\sigma}+\frac{1}{m}\Big)+m_cL_f\bar{a}\frac{1}{m-\sigma}<1.$
\end{itemize}

\begin{theorem}\label{theorem1}
	The system (\ref{cohen1}) admits a unique exponentially stable Poisson stable solution provided that conditions $(C1)-(C9)$ are fulfilled.
\end{theorem}
\noindent {\bf Proof.} Let us show the completeness of the set $\cal B.$ Consider a sequence $\phi^k(t)$ in $\cal B,$ which converges on $\mathbb{R}$ to  a limit  function $\phi(t).$ Fix a section $I\subset \mathbb{R}.$ We have that
\begin{eqnarray}\label{Poisson}
	\| \phi(t+t_p)-\phi(t)\| \le \| \phi(t+t_p)-\phi^k(t+t_p)\| +\| \phi^k(t+t_p)-\phi^k(t)\| +\| \phi^k(t)-\phi(t)\|. 	
\end{eqnarray}

One can take sufficiently large numbers $p$ and $k$ such that each term on the right-hand-side of (\ref{Poisson}) is smaller than $\frac{\epsilon}{3}$ for an arbitrary $\epsilon>0$ and $t\in I$. The inequality (\ref{Poisson})  implies that $\phi(t+t_p)$ converges to $\phi(t)$ uniformly on $I.$ That is, the set  $\cal B$ is complete. 

By Lemmas \ref{lemma1} and \ref{lemma3}, on the invariance and contractiveness of the operator $T$ in the set $\cal B,$  one can obtain that there  exists a unique fixed point $z \in {\cal B}$ of the operator $T,$ which is a solution of  the system (\ref{cohen3}) and satisfies the convergence property. Thus, the function $z(t)=(z_1(t),z_2(t),\ldots,z_n(t))$ is a unique Poisson stable solution of the system (\ref{cohen3}).	

Now, let us discuss the stability of the  solution $z(t).$ 

It is true that the solution $z(t)=(z_1(t),z_2(t),\ldots,z_n(t))$ satisfies the integral inequality
\begin{eqnarray*}
	z_{i}(t)=z_i(t_0)e^{-\int_{t_0}^{t}u_{i}(z_i(\tau))d\tau}+\int_{t_0}^{t} e^{-\int_{s}^{t}u_{i}(z_i(\tau))d\tau}\Big(\sum_{j=1}^{n}c_{ij}(s)f_j(h^{-1}_{j}(z_j(s)))+v_{i}(s)\Big)ds,
\end{eqnarray*}
for all $ i=1, \ldots, n.$

Let $y(t)=(y_{1}(t),y_2(t),\dots,y_n(t)),$ $ i=1, \ldots, n,$ be another solution of system  (\ref{cohen3}). Then, for each $ i=1, \ldots, n,$ we have that
\begin{eqnarray*}
	y_{i}(t)=y_i(t_0)e^{-\int_{t_0}^{t}u_{i}(y_i(\tau))d\tau}+\int_{t_0}^{t} e^{-\int_{s}^{t}u_{i}(y_i(\tau))d\tau}\Big(\sum_{j=1}^{n}c_{ij}(s)f_j(h^{-1}_{j}(y_j(s)))+v_{i}(s)\Big)ds,
\end{eqnarray*}
and 
\begin{eqnarray} \label{dif}
	&& y_{i}(t)-z_{i}(t) = (y_i(t_0)-z_i(t_0))e^{-\int_{t_0}^{t}u_{i}(z_i(\tau))d\tau}+y_i(t_0)(e^{-\int_{t_0}^{t}u_{i}(y_i(\tau))d\tau}-e^{-\int_{t_0}^{t}u_{i}(z_i(\tau))d\tau})+\nonumber\\
	&&\int_{t_0}^{t} (e^{-\int_{s}^{t}u_{i}(y_i(\tau))d\tau}-e^{-\int_{t_0}^{t}u_{i}(z_i(\tau))d\tau})\Big(\sum_{j=1}^{n}c_{ij}(s)f_j(h^{-1}_{j}(y_j(s)))+v_i(s)\Big)ds-\nonumber\\
	&&\int_{t_0}^{t} e^{-\int_{s}^{t}u_{i}(z_i(\tau))d\tau}\Big(\sum_{j=1}^{n}c_{ij}(s)f_j(h^{-1}_{j}(y_j(s)))-\sum_{j=1}^{n}c_{ij}(s)f_j(h^{-1}_{j}(z_j(s)))\Big)ds.
\end{eqnarray}
Denote $\omega(t)=y(t)-z(t)$ and $\omega(t_0)=y(t_0)-z(t_0),$ where $\omega(t)=(\omega_1(t),\omega_2(t),\ldots,\omega_n(t)).$ Relation (\ref{dif}) implies that $\omega(t)$ satisfy the next equation,
\begin{eqnarray} \label{dif3}
	&& \omega_{i}(t) = \omega_i(t_0))e^{-\int_{t_0}^{t}u_{i}(z_i(\tau))d\tau}+(\omega_i(t_0)+z_i(t_0))(e^{-\int_{t_0}^{t}u_{i}(\omega_i(\tau)+z_i(\tau))d\tau}-e^{-\int_{t_0}^{t}u_{i}(z_i(\tau))d\tau})+\nonumber\\
	&&\int_{t_0}^{t} (e^{-\int_{s}^{t}u_{i}(\omega_i(\tau)+z_i(\tau))d\tau}-e^{-\int_{t_0}^{t}u_{i}(z_i(\tau))d\tau})\Big(\sum_{j=1}^{n}c_{ij}(s)f_j(h^{-1}_{j}(\omega_j(s)+z_j(s)))+v_i(s)\Big)ds-\nonumber\\
	&&\int_{t_0}^{t} e^{-\int_{s}^{t}u_{i}(z_i(\tau))d\tau}\Big(\sum_{j=1}^{n}c_{ij}(s)f_j(h^{-1}_{j}(\omega_j(s)+z_j(s)))-\sum_{j=1}^{n}c_{ij}(s)f_j(h^{-1}_{j}(z_j(s)))\Big)ds,
\end{eqnarray}
for $ i=1, \ldots, n.$

Now, let us construct the sequence of successive approximation $\omega^k(t),$ $k\geq 0,$ such that
\begin{eqnarray}\label{omega}
	\omega^{0}(t)=(\omega_1(t_0)e^{-\int_{t_0}^{t}u_{1}(z_1(\tau))d\tau},\omega_2(t_0)e^{-\int_{t_0}^{t}u_{2}(z_2(\tau))d\tau},\cdots,\omega_n(t_0)e^{-\int_{t_0}^{t}u_{n}(z_n(\tau))d\tau}).
\end{eqnarray}

In what follows, inequality
\begin{eqnarray}\label{expp}
	|e^{-\int_{t_0}^{t}u_{i}(y_i(\tau))d\tau}-e^{-\int_{t_0}^{t}u_{i}(z_i(\tau))d\tau}|\le e^{-m_i(t-t_0)}\int_{t_0}^{t}L^u_i|y_i(\tau)-z_i(\tau)|d \tau, t\geq t_0,   i=1,2,\cdots,n,
\end{eqnarray}
will be intensively utilized.
To approve (\ref{expp}), formula $e^x-e^y=e^c(x-y)$ is applied with $x=-\int_{t_0}^{t}u_{i}(y_i(\tau))d\tau, y=-\int_{t_0}^{t}u_{i}(z_i(\tau))d\tau,$ and a number $c$ between $x$ and $y.$ Moreover, it is easy to check that  relation $c<-m_i(t-t_0)$ is correct.

Therefore, using (\ref{dif3}), we obtain that for each $i=1,2,\ldots,n,$ the following inequalities are correct,
\begin{eqnarray} \label{diff}
	&& |\omega^{k+1}_{i}(t)| \le|\omega_i(t_0)|e^{-m_i(t-t_0)}+He^{-m_i(t-t_0)}\int_{t_0}^{t}L^u_i|\omega^{k}_i(\tau)|d \tau+\nonumber \\
	&&\int_{t_0}^{t}e^{-m_i(t-t_0)}\Big(\int_{t_0}^{s}L^u_i|\omega^{k}_i(\tau)|d \tau\Big)(\sum_{j=1}^{n}m^c_{ij}m^f_i+m^v_i)ds+\int_{t_0}^{t}e^{-m_i(t-t_0)}\sum_{j=1}^{n}m^c_{ij}L^f_i\bar{a_i}|\omega^{k}_i(s)|ds. 
\end{eqnarray}
From (\ref{omega}) it follows  that
\begin{eqnarray*}	&&\|\omega^0(t)\|\le (\|\omega(t_0)\|+\epsilon)e^{-\sigma(t-t_0)},\end{eqnarray*}
where $\epsilon$ is a positive number such that
\begin{eqnarray}\label{epsilon}
	\epsilon>\|\omega(t_0)\|\frac{HL_u\frac{1}{\sigma}+L_u(m_cm_f+m_v)\frac{1}{\sigma}\Big(\frac{2}{m-\sigma}+\frac{1}{m}\Big)+m_cL_f\bar{a}\frac{1}{m-\sigma}}{1-HL_u\frac{1}{\sigma}-L_u(m_cm_f+m_v)\frac{1}{\sigma}\Big(\frac{2}{m-\sigma}+\frac{1}{m}\Big)-m_cL_f\bar{a}\frac{1}{m-\sigma}}.
\end{eqnarray}

Assume that for fixed $k \in \mathbb{N}$ the following inequality is valid:
\begin{eqnarray*}
	\|\omega^{k}(t)\|\le(\|\omega(t_0)\|+\epsilon)e^{-\sigma(t-t_0)}.
\end{eqnarray*}
Applying inequality (\ref{diff}) we get that
\begin{eqnarray*}
	&&\|\omega^{k+1}(t)\|\le\|\omega(t_0)\|e^{-m(t-t_0)}+He^{-m(t-t_0)}\int_{t_0}^{t}L_u(\|\omega(t_0)\|+\epsilon)e^{-\sigma(s-t_0)}ds+\nonumber\\
	&&\int_{t_0}^{t}e^{-m(t-s)}L_u(m_cm_f+m_v)\Big[\int_{t_0}^{s}(\|\omega(t_0)\|+\epsilon)e^{-\sigma(\tau-t_0)} d \tau\Big]ds+\int_{t_0}^{t}e^{-m(t-s)}m_cL_f\bar{a}(\|\omega(t_0)\|+\epsilon)e^{-\sigma(s-t_0)}ds\leq \\
	&&\|\omega(t_0)\|e^{-m(t-t_0)}+He^{-m(t-t_0)}L_u(\|\omega(t_0)\|+\epsilon)\frac{1}{\sigma}|e^{-\sigma(t-t_0)}-1|+\\
	&&\int_{t_0}^{t}e^{-m(t-s)}L_u(m_cm_f+m_v)(\|\omega(t_0)\|+\epsilon)\frac{1}{\sigma}|e^{-\sigma(s-t_0)}-1|ds+\int_{t_0}^{t}e^{-m(t-s)}m_cL_f\bar{a}(\|\omega(t_0)\|+\epsilon)e^{-\sigma(s-t_0)}ds\leq \\
	&&\|\omega(t_0)\|e^{-m(t-t_0)}+He^{-m(t-t_0)}L_u(\|\omega(t_0)\|+\epsilon)\frac{1}{\sigma}|e^{-\sigma(t-t_0)}-1|+\\
	&&L_u(m_cm_f+m_v)(\|\omega(t_0)\|+\epsilon)\frac{1}{\sigma}\Big(\frac{1}{m-\sigma}[e^{-\sigma(t-t_0)}+e^{-m(t-t_0)}]+\frac{1}{m}e^{-m(t-t_0)}\Big)+\\
	&&m_cL_f\bar{a}(\|\omega(t_0)\|+\epsilon)\frac{1}{m-\sigma}[e^{-\sigma(t-t_0)}-e^{-m(t-t_0)}]\le \|\omega(t_0)\|e^{-\sigma(t-t_0)}+He^{-\sigma(t-t_0)}L_u(\|\omega(t_0)\|+\epsilon)\frac{1}{\sigma}+\\
	&&L_u(m_cm_f+m_v)(\|\omega(t_0)\|+\epsilon)\frac{1}{\sigma}\Big(\frac{2}{m-\sigma}+\frac{1}{m}\Big)e^{-\sigma(t-t_0)}+m_cL_f\bar{a}(\|\omega(t_0)\|+\epsilon)\frac{1}{m-\sigma}e^{-\sigma(t-t_0)}\le \\
	&&\Big(\|\omega(t_0)\|+\Big(HL_u\frac{1}{\sigma}+L_u(m_cm_f+m_v)\frac{1}{\sigma}\Big(\frac{2}{m-\sigma}+\frac{1}{m}\Big)+m_cL_f\bar{a}\frac{1}{m-\sigma}\Big)(\|\omega(t_0)\|+\epsilon)\Big)e^{-\sigma(t-t_0)},
\end{eqnarray*}
for all $k=0,1,\cdots.$ Condition (C9) and assumption (\ref{epsilon}) imply that $\|\omega^{k}(t)\|\le(\|\omega(t_0)\|+\epsilon)e^{-\sigma(t-t_0)}$ for each $k=0,1,\cdots.$  

Now, let us show that the sequence $\omega^{k}(t)$ uniformly converges. Applying inequality (\ref{diff}), we obtain that

\begin{eqnarray*}
	&&	\|\omega^1(t)-\omega^0(t)\|\le He^{-m(t-t_0)}\int_{t_0}^{t}L_u\|\omega^{0}(\tau)\|d \tau+\int_{t_0}^{t}e^{-m(t-t_0)}\Big(\int_{t_0}^{s}L_u\|\omega^{0}(\tau)\|d \tau\Big)(m_cm_f+m_v)ds+\nonumber \\
	&&\int_{t_0}^{t}e^{-m(t-t_0)}m_cL_f\bar{a}\|\omega^{0}(s)\|ds\le HL_ue^{-m(t-t_0)}\int_{t_0}^{t}(\|\omega(t_0)\|+\epsilon)e^{-\sigma(s-t_0)}ds+\\
	&&\int_{t_0}^{t}e^{-m(t-s)}L_u(m_cm_f+m_v)\Big[\int_{t_0}^{s}(\|\omega(t_0)\|+\epsilon)e^{-\sigma(\tau-t_0)} d \tau\Big]ds+\int_{t_0}^{t}e^{-m(t-s)}m_cL_f\bar{a}(\|\omega(t_0)\|+\epsilon)e^{-\sigma(s-t_0)}ds\leq \\
	&&\Big(HL_u\frac{1}{\sigma}+L_u(m_cm_f+m_v)\frac{1}{\sigma}\Big(\frac{2}{m-\sigma}+\frac{1}{m}\Big)+m_cL_f\bar{a}\frac{1}{m-\sigma}\Big)(\|\omega(t_0)\|+\epsilon)e^{-\sigma(t-t_0)},
\end{eqnarray*}
and
\begin{eqnarray*}
	&&\|\omega^2(t)-\omega^1(t)\|\le He^{-m(t-t_0)}\int_{t_0}^{t}L_u\|\omega^1(\tau)-\omega^0(\tau)\|d\tau+\\
	&&\int_{t_0}^{t}e^{-m(t-s)}\Big(\int_{t_0}^{t}L_u\|\omega^1(\tau)-\omega^0(\tau)\|d\tau\Big)(m_cm_f+m_v)ds+\int_{t_0}^{t}e^{-m(t-t_0)}m_cL_f\bar{a}\|\omega^1(s)-\omega^0(s)\|ds \le\\
	&&\Big(HL_u\frac{1}{\sigma}+L_u(m_cm_f+m_v)\frac{1}{\sigma}\Big(\frac{2}{m-\sigma}+\frac{1}{m}\Big)+m_cL_f\bar{a}\frac{1}{m-\sigma}\Big)^2(\|\omega(t_0)\|+\epsilon)e^{-\sigma(t-t_0)}.
\end{eqnarray*}

By the method of mathematical induction, it can be shown that

\begin{eqnarray*}
	&&\|\omega^{k+1}(t)-\omega^k(t)\|\le \Biggl(HL_u\frac{1}{\sigma}+L_u(m_cm_f+m_v)\frac{1}{\sigma}\Big(\frac{2}{m-\sigma}+\frac{1}{m}\Big)+m_cL_f\bar{a}\frac{1}{m-\sigma}\Biggl)^{k+1}(\|\omega(t_0)\|+\epsilon)e^{-\sigma(t-t_0)},
\end{eqnarray*}
for all $k\geq 0.$ Condition (C9) gives that $\sup_{t \in [t_0,\infty)}\|\omega^{k+1}(t)-\omega^k(t)\|\rightarrow 0$ as $k \rightarrow \infty.$ Thus, the sequence $\omega^k(t)$ uniformly converges to the unique solution, $\omega(t)=y(t)-z(t),$ of the integral equation (\ref{dif3}), which satisfy inequality
\begin{eqnarray}
	\|y(t)-z(t)\|\le(\|y(t_0)-z(t_0)\|+\epsilon)e^{-\sigma(t-t_0)}.
\end{eqnarray}
Consequently, the Poisson stable solution $z(t)$ of the system (\ref{cohen3}) is exponentially stable. 

Now, consider a function $w(t)=(w_1(t),w_2(t),\ldots,w_n(t)),$ such that $w_i(t)=h^{-1}_i(z_i(t)),$ $i=1,2,\ldots,n.$ According  to substitution (\ref{subst}), the function $w(t)$ is a unique solution of system (\ref{cohen1}). let us show that $w(t)$ is Poisson stable. Using inequality (\ref{h2}), on a fixed bounded interval $I\subset \mathbb{R},$ we obtain that
\begin{eqnarray*}
	&&|w_i(t+t_p)-w_i(t)|=|h^{-1}_i(z_i(t+t_p))-h^{-1}_i(z_i(t))|\le\bar{a_i}|z_i(t+t_p)-z_i(t)|,
\end{eqnarray*}
for all $i=1,2,\ldots,n.$ Thus, each sequence $w_i(t+t_p), i=1,2,\ldots,n,$ uniformly converges to $w_i(t),$ $t\in I,$ as $p\rightarrow \infty,$ and one can conclude that the function $w(t)=(w_1(t),w_2(t),\ldots,w_n(t))$ is a unique Poisson stable solution of neural network (\ref{cohen1}). 

Finally, we will check that the solution $w(t)$ is exponentially stable. If $x(t)=h^{-1}(y(t))$ is another solution of system (\ref{cohen1}) then we obtain that 
\begin{eqnarray*}
	&&\|x(t)-w(t)\|=\|h^{-1}(y(t))-h^{-1}(z(t))|\le \bar{a}\|y(t)-z(t)\|\le \bar{a}(\|y(t_0)-z(t_0)\|+\epsilon)e^{-\sigma(t-t_0)}.
\end{eqnarray*}
Consequently, the Poisson stable solution $w(t)$ of the neural network (\ref{cohen1}) is exponentially stable. $\Box$

\begin{theorem}\label{theorem2} Assume that conditions $(C1)-(C9)$ are satisfied. Then CGNN (\ref{cohen1}) admits a unique exponentially stable unpredictable solution.
\end{theorem}
\noindent {\bf Proof.} Due to the previous theorem, there exists a unique exponentially stable Poisson stable solution $w(t)=h^{-1}(z(t))$ of neural network (\ref{cohen1}). Now, we will prove the unpredictability of $w(t).$ Firstly, we will show that the Poisson stable solution $z(t)$ of system (\ref{cohen3}) satisfies the separation property.

Applying the relations
\begin{eqnarray*}
	&&z_i(t)=z_{i}(s_p)-\int_{s_p}^{t}u_i(z_i(s))z_i(s)ds+\int_{s_p}^{t}\sum_{j=1}^{n}c_{ij}(s)f_j(h^{-1}_j(z_j(s)))ds+\int_{s_p}^{t}v_i(s)ds
\end{eqnarray*} 
and
\begin{eqnarray*}
	z_i(t+t_p)=z_{i}(t+t_p)-\int_{s_p}^{t}u_i(z_i(s+t_p))z_i(s+t_p)ds+\int_{s_p}^{t}\sum_{j=1}^{n}c_{ij}(s+t_p)f_j(h^{-1}_j(z_j(s+t_p)))ds+\int_{s_p}^{t}v_i(s+t_p)ds,
\end{eqnarray*}
we obtain that
\begin{eqnarray*}
	&&z_i(t+t_p)-z_i(t)=z_{i}(t+t_p)-z_{i}(t_p)-\int_{s_p}^{t}u_i(z_i(s+t_p))z_i(s+t_p)ds+\int_{s_p}^{t}u_i(z_i(s))z_i(s)ds+\\
	&&\int_{s_p}^{t}\sum_{j=1}^{n}c_{ij}(s+t_p)f_j(h^{-1}_j(z_j(s+t_p)))ds-\int_{s_p}^{t}\sum_{j=1}^{n}c_{ij}(s)f_j(h^{-1}_j(z_j(s)))ds+\int_{s_p}^{t}v_i(s+t_p)ds-\int_{s_p}^{t}v_i(s)ds,
\end{eqnarray*}
for each $i=1,2,\dots,n.$

There exist positive numbers $\delta_1$ and integers $l,k$ such that, for each $i,j=1,2,\dots,n,$  the following inequalities are satisfied:
\begin{equation}\label{1}
	\delta_1 <\delta;
\end{equation}
\begin{equation}\label{2}
	|c_{ij}(t+s)-c_{ij}(s)|<\epsilon_0 (\frac{1}{l}+\frac{2}{k}), \quad t\in \mathbb{R}, 
\end{equation}
\begin{equation}\label{3}
	|v_i(t+s)-v_i(s)|<\epsilon_0 (\frac{1}{l}+\frac{2}{k}), \quad t\in \mathbb{R},
\end{equation}
\begin{equation}\label{4}
	\delta_1 \Big(1- (\frac{1}{l}+\frac{2}{k})(L^u_iH+M_i+n m^f_i+\sum_{j=1}^{n} m^c_{ij}L^f_i\bar{a_i})\Big)>\frac{3}{2l}, 
\end{equation}
\begin{equation}\label{5}
	|z_{i}(t+s)-z_{i}(t)|<\epsilon_0 \min (\frac{1}{k},  \frac{1}{4l}),  \quad t\in \mathbb{R}, |s|<\delta_1.
\end{equation}
Let the numbers  $\delta_1, l$ and $k$, as well as numbers $n \in  \mathbb{N},$ and $i=1,\dots,n$, be fixed. Consider the following two alternatives: (i)  $|z_{i}(t_p+s_p)-z_{i}(s_p)| <\epsilon_0/l;$ \quad (ii) $|z_{i}(t_p+s_p)-z_{i}(s_p)| \geq \epsilon_0/l.$  

(i) Using (\ref{5}), one can show that 
\begin{eqnarray}\label{omeg}
	&&|z_{i}(t+t_p)-z_{i}(t_p)|\leq
	|z_{i}(t+t_p)-z_{i}(t_p+s_p)|+ |z_{i}(t_p+s_p)-z_{i}(s_p)|\nonumber + |z_{i}(s_p)-z_{i}(t)|\\&&< \frac{\epsilon_0}{l}+\frac{\epsilon_0}{k}+\frac{\epsilon_0}{k} =\epsilon_0 (\frac{1}{l}+\frac{2}{k}), \ i=1,2,\cdots,n,
\end{eqnarray}
if $t \in [s_p, s_p+\delta_1].$ The inequalities  (\ref{1})--(\ref{omeg}) imply that

\begin{eqnarray*}
	&&|z_i(t+t_p)-z_i(t)|\geq \int_{s_p}^{t}|v_i(s+t_p)-v_i(s)|ds-|z_{i}(t+t_p)-z_{i}(t_p)|-\\
	&&\int_{s_p}^{t}\Big(|u_i(z_i(s+t_p))-u_i(z_i(s))||z_i(s+t_p)|+|u_i(z_i(s))||z_i(s+t_p)-z_i(s)|\Big)ds-\\
	&&\int_{s_p}^{t}\sum_{j=1}^{n}\Big(|c_{ij}(s+t_p)-c_{ij}(s)||f_j(h^{-1}_j(z_(s+t_p)))|+\sum_{j=1}^{n}|c_{ij}(s)||f_j(h^{-1}_j(z_(s+t_p)))-f_j(h^{-1}_j(z_(s)))|\Big)ds >\\
	&&\delta_1 \epsilon_0-\frac{\epsilon_0}{l}-\delta_1(L^u_iH+M_i) \epsilon_0 (\frac{1}{l}+\frac{2}{k})- \delta_1 \Big(n \epsilon_0 (\frac{1}{l}+\frac{2}{k}) m^f_i+\sum_{j=1}^{n} m^c_{ij}L^f_i\bar{a_i}\epsilon_0 (\frac{1}{l}+\frac{2}{k})\Big)>\frac{\epsilon_0}{2l}
\end{eqnarray*}
for $t \in[s_p,s_p+\delta_1].$

(ii) If $|z_{i}(t_p+s_p)-z_{i}(s_p)| \geq \epsilon_0/l$, it is not difficult to find that  (\ref{5}) implies:
\begin{eqnarray*}
	&&|z_{i}(t+t_p)-z_{i}(t)|\geq  |z_{i}(t_p+s_p)-z_{i}(s_p)| - |z_{i}(s_p)-z_{i}(t)|- \nonumber|z_i(t+t_p)-z_i(t_p+s_p)|>\\&&  \frac{\epsilon_0}{l}-\frac{\epsilon_0}{4l}-\frac{\epsilon_0}{4l}=\frac{\epsilon_0}{2l}, \ i=1,2,\dots,n,
\end{eqnarray*}
if  $t \in [s_p-\delta_1, s_p+\delta_1]$  and $p \in \mathbb N.$ So, the cases $(i)$ and $(ii)$ imply that the solution $z(t)$ satisfies separation property, and it can be concluded that $z(t)$ is an unpredictable solution of system (\ref{cohen3}), with sequences $t_p,$ $s_p$ and positive numbers $\frac{\delta_1}{2},$ $\frac{\epsilon_0}{2l}.$

Finally, we show that the solution $w(t)=h^{-1}(z(t))$ of the neural network (\ref{cohen1}) is also unpredictable. Actually, applying condition (\ref{h2}), we get that
\begin{eqnarray*}
	&&|w_i(t+t_p)-w_i(t)|=|h^{-1}_i(z_i(t+t_p))-h^{-1}_i(z_i(t))|\geq\underline{a}_i|z_i(t+t_p)-z_i(t)|>\underline{a}_i\frac{\epsilon}{2l}, \quad i=1,2,\cdots,n,
\end{eqnarray*}
for all $t\in[s_p-\frac{\delta_1}{2}, s_p+\frac{\delta_1}{2}].$ Thus, the neural network (\ref{cohen1}) admits a unique exponentially stable unpredictable solution. $\Box$
\section{The model with compartmental strength of connectivity and inputs} \label{compart1}
In order to increase  the applicability of this study, CGNNs (\ref{cohen1}) with compartmental periodic unpredictable strengths of connectivity between cells , $c_{ij}(t),$ and input data, $v_i(t),$  are considered. Under additional conditions, Theorems \ref{theorem3} and \ref{theorem4} on the Poisson stability and unpredictability in the neural networks are proved, in this section.

 Assume that the following condition is valid,
	\begin{itemize}
	\item[(C10)] functions $c_{ij}(t)$  and $v_{i}(t),$ $i=1,2,\cdots,n,$ $j=1,2,\cdots,n,$ are compartmental periodic unpredictable such that
	 $c_{ij}(t)=C_{ij}(t,t),$  $v_{i}(t)=V_{i}(t,t),$ where the functions $C_{ij}(\theta,\tau),$  and $V_{i}(\theta,\tau)$ are $\omega-$periodic in $\theta$ uniformly with respect to $\tau,$ and unpredictable in $\tau$ with common sequences  of convergence $t_p,$  and separation $s_p,$ $p=1,2,\ldots,$ uniformly with respect to $\theta.$
\end{itemize}
Denote $\displaystyle \bar{m}^{c}_{ij}= \sup_{\theta,\tau \in \mathbb{R}}|C_{ij}(\theta,\tau)|,$ $\bar{m}^{v}_{i}= \sup_{\theta,\tau \in \mathbb{R}}|V_{i}(\theta,\tau)|.$

The following assumptions are required,
	\begin{itemize}
	\item[(C11)] the convergence sequence $t_p$ satisfies kappa property with respect to $\omega;$
	\item[(C12)] $\displaystyle \frac{1}{m_i}\Big(\sum_{j=1}^{n}\bar{m}^{c}_{ij}m^f_i+\bar{m}^{v}_{i}\Big)<H;$ 
	\item[(C13)] $\displaystyle \frac{1}{m_i}\Big(\frac{L^u_i}{m_i}(\sum_{j=1}^{n}\bar{m}^{c}_{ij}m^f_i+\bar{m}^{v}_{i})+\sum_{j=1}^{n}\bar{m}^{c}_{ij}L^f_i\bar{a_i}\Big)<1;$
	\item[(C14)]	$\displaystyle HL_u\frac{1}{\sigma}+L_u(\bar{m_c}m_f+\bar{m_v})\frac{1}{\sigma}\Big(\frac{2}{m-\sigma}+\frac{1}{m}\Big)+\bar{m_c}L_f\bar{a}\frac{1}{m-\sigma}<1,$
\end{itemize}
where $\displaystyle \bar{m_c}=\max_{(i)} \sum_{j=1}^{n}\bar{m}^c_{ij},$  $\displaystyle \bar{m_v}=\max_{(i)} \bar{m}^v_i, i=1,2,\ldots,n.$

\begin{theorem}\label{theorem3} Let conditions $(C1)-(C3),$ $(C5),$ $(C6),$ and $(C10)-(C14)$ are valid. Then neural network (\ref{cohen1}) possesses a unique exponentially stable Poisson stable solution.
\end{theorem}
\noindent {\bf Proof.} Under conditions $(C10),$ $(C11)$ and Theorem \ref{lemma333}, the functions $c_{ij}(t)$ and  $v_{i}(t)$ are unpredictable. Therefore, using the technique of proving Theorem \ref{theorem1}, one can ensure that the neural network (\ref{cohen1}) admits a unique Poisson stable solution with exponential property. $\Box$

Similarly to the proofs of Theorems \ref{theorem1} and \ref{theorem2}, it can be shown that the following statement is true.
\begin{theorem}\label{theorem4} Assume that conditions $(C1)-(C3),$ $(C5),$ $(C6),$ and $(C10)-(C14)$ are satisfied. Then neural network (\ref{cohen1}) possesses a unique exponentially stable unpredictable solution.
\end{theorem}

\section{Numerical analysis} \label{numexmple}

This part of the paper contains three examples of the neural networks, which approve the theoretical results  of the main body. They demonstrate  the chaotic nature of the dynamics in all three models, and two last ones are simulated with compartmental strengths of connectivity and inputs to find that a special parametric characteristic, degree of periodicity, can be utilized for estimation of contributions of components such as periodicity and unpredictability.  

For shaping unpredictable inputs of the neural networks, we shall use results of previous papers, in particular \cite{Compart},  such that the functions are products of hybrid systems.   Let us take into account the logistic mapping equation
\begin{eqnarray*}
	\lambda_{i+1}=\mu\lambda_i(1-\lambda_i),
\end{eqnarray*} 
where $i\in \mathbb{Z},$ $\mu \in [3+(2/3)^{1/2},4].$ Define a piecewise continuous function $\pi(t)$ such that $\pi(t)=\lambda_i \xi(t-ih),$ $t\in(ih, (i+1)h],$ where $h$ is a positive constant, and $\xi(t):(0,h] \rightarrow\mathbb{R}$ is a continuous function. In paper \cite{Compart}, it was proved that the function $\pi(t)$ is discontinuous unpredictable function. Moreover, applying the function $\pi(t),$ it was constructed continuous unpredictable function $\Xi(t)=\int_{-\infty}^{t}e^{-\alpha(t-s)}\pi(s)ds,$ where $\alpha$ is a positive real number. 

The number $h$ is said to be \textit{the length of step}  
of the functions $\pi(t)$ and $\Xi(t).$ For compartmental unpredictable functions, the ratio of the period and the length of step, $\nabla=\omega/h,$ is called \textit{the degree of periodicity}.

Below we will use unpredictable function $\Theta(t)=\int_{-\infty}^{t}e^{-3(t-s)}\pi(s)ds,$ with  $\pi(t)=\lambda_i (t-i)$ if $t\in(ih, (i+1)h].$ 

Firstly, let us show the dynamics of CGNNs (\ref{cohen1}) with unpredictable synaptic connections and inputs.

\noindent \textbf{Example 1.}  Let us consider of the following CGNNs,
\begin{eqnarray}\label{ex1}
	&&x'_{i}(t)=-a_{i}(x_{i}(t))\Big[b_i(x_i(t))-\sum_{j=1}^{n} c_{ij}(t)f_j(x_{j}(t)) + v_{i}(t)\Big],
\end{eqnarray}
where $i=1,2,3,$ $f(s)=0.5\arctan(s),$ $a_1(s)=\cos(0.1s),$ $a_2(s)=\cos(0.2s),$ $a_3(s)=\cos(0.05s),$ $b_1(s)=2\sin(0.4s),$ $b_2(s)=6\sin(0.5s),$ $b_3(s)=4\sin(0.2s),$
$c_{11}(t)=0.2\Theta(t),$ $c_{12}(t)=0.05\Theta(t),$ $c_{13}(t)=0.1\Theta(t),$ $c_{21}(t)=0.1\Theta(t),$ $c_{22}(t)=0.2\Theta(t),$ $c_{23}(t)=0.05\Theta(t),$  $c_{31}(t)=0.05\Theta(t),$ $c_{32}(t)=0.1\Theta(t),$ $c_{33}(t)=0.2\Theta(t),$ $v_1(t)=1.5\Theta(t),$ $v_2(t)=2.5\Theta(t),$ $v_3(t)=2\Theta(t).$ 
Calculate $H_0=1,$ $\bar{a_1}=0.995,$ $\bar{a_2}=0.98,$ $\bar{a_3}=0.999,$ $m^f_1=m^f_2=m^f_3=\pi/4,$ $L^f_1=L^f_2=L^f_3=0.5,$ $\sum_{j=1}^{3}m^c_{1j}=\sum_{j=1}^{3}m^c_{2j}=\sum_{j=1}^{n}m^c_{3j}=7/60,$  $m^v_{1}=0.5,$ $m^v_{2}=0.84,$ and $m^v_{3}=0.67.$ 

\begin{figure}[ht]
	\centering
	\includegraphics[width=9cm]{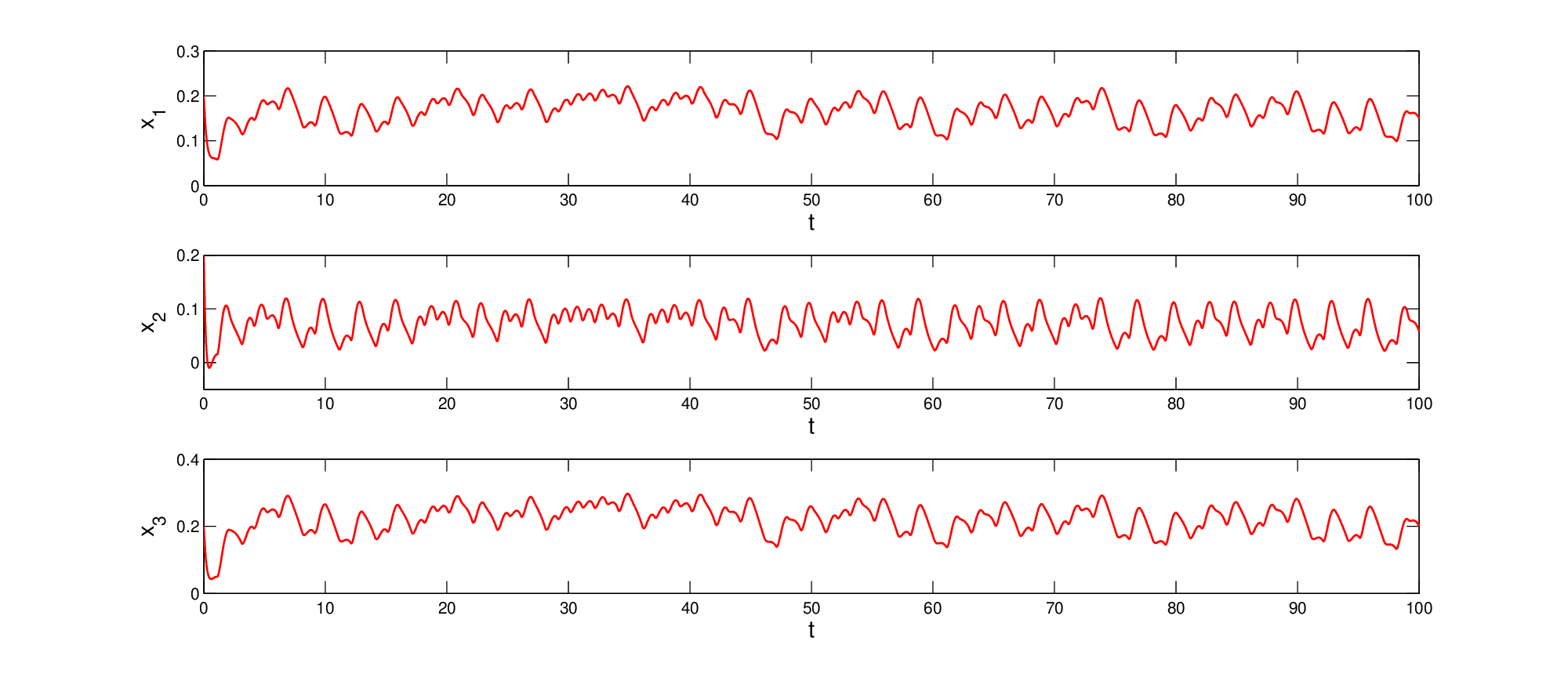}
	\caption{The coordinates of solution $x(t)$ of the neural network (\ref{ex1}) with initial values $x_1(0)=0.2, x_2(0)=0.2,$ and  $x_3(0)=0.2.$}  \label{fig01}
\end{figure}

\begin{figure}[ht]
	\centering
	\includegraphics[width=9cm]{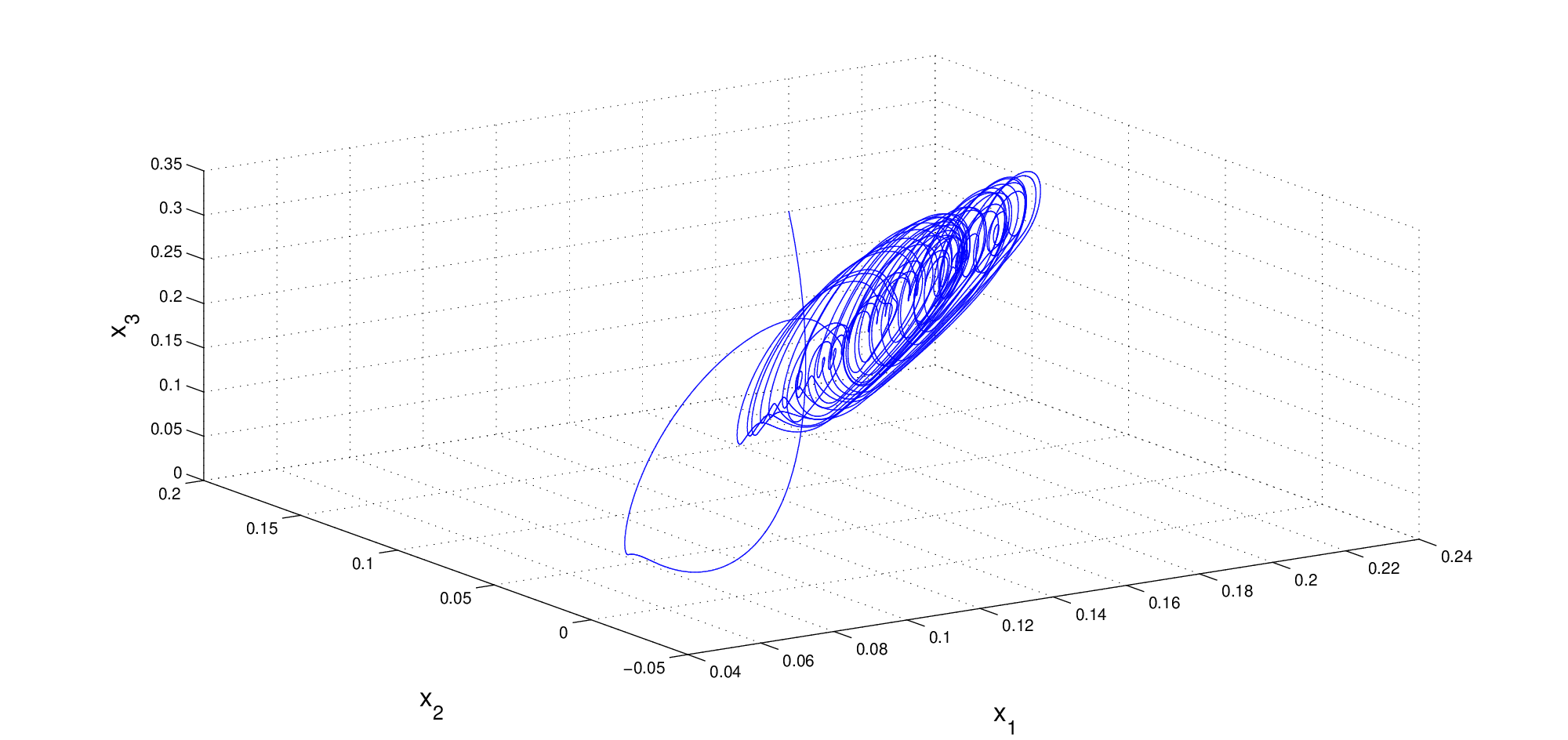}
	\caption{Trajectory of the solution $x(t).$} \label{fig02}
\end{figure}

Taking into account that $h_i(s)=\int_{0}^{s}\frac{1}{a_i(\tau)}d\tau,$ $i=1,2,3,$ one can find $h^{-1}_1(s)=\int_{0}^{s}\cos(0.1\tau)d\tau=10\sin(0.1s),$ $h^{-1}_2(s)=\int_{0}^{s}\cos(0.2\tau)d\tau=5\sin(0.2s),$ $h^{-1}_3(s)=\int_{0}^{s}\cos(0.05\tau)d\tau=20\sin(0.05s).$ The functions $u_i(s)=\frac{b_i(h^{-1}_i(s))}{s},$ $i=1,2,3,$ such that $u_1(s)=\frac{2\sin(2(\sin(0.1s)))}{s},$ $u_2(s)=\frac{6\sin(2.5(\sin(0.2s)))}{s},$ $u_3(s)=\frac{4\sin(4(\sin(0.05s)))}{s}.$ Thus, we get that $0.78\le m_1\le 2;$ $2.86\le m_2\le 6;$ $0.8\le m_3\le 4,$ and the function $u_i(s),$ $i=1,2,3,$ satisfy Lipschitz condition with $L^u_1=0.02,$ $L^u_2=0.06,$ $L^u_3=0.04.$ Conditions (C1)--(C8) are satisfied with above described functions and constants. The assumption (C9) is valid since $H=1.1,$ $L_u=0.06,$ $\sigma=0.4,$ $m=0.78,$ $m_c=7/60,$ $m_f=\pi/4,$ $L_f=0.5$ and $\bar{a}=1.$ According Theorem \ref{theorem2} there exists a unique unpredictable solution, $z(t)=(z_{1}(t),z_{2}(t),z_{3}(t)),$ of neural network (\ref{ex1}). In Figures \ref{fig01} and \ref{fig02} the solution $x(t)=(x_{1}(t),x_{2}(t),x_{3}(t)),$ which exponentially 
converges to the unpredictable solution $w(t)$ is shown.

In the next examples we consider CGNNs (\ref{ex1}) with compartmental periodic unpredictable inputs, and show how its dynamics depends on the degree of periodicity.

\begin{figure}[ht]
	\centering
	\includegraphics[width=9cm]{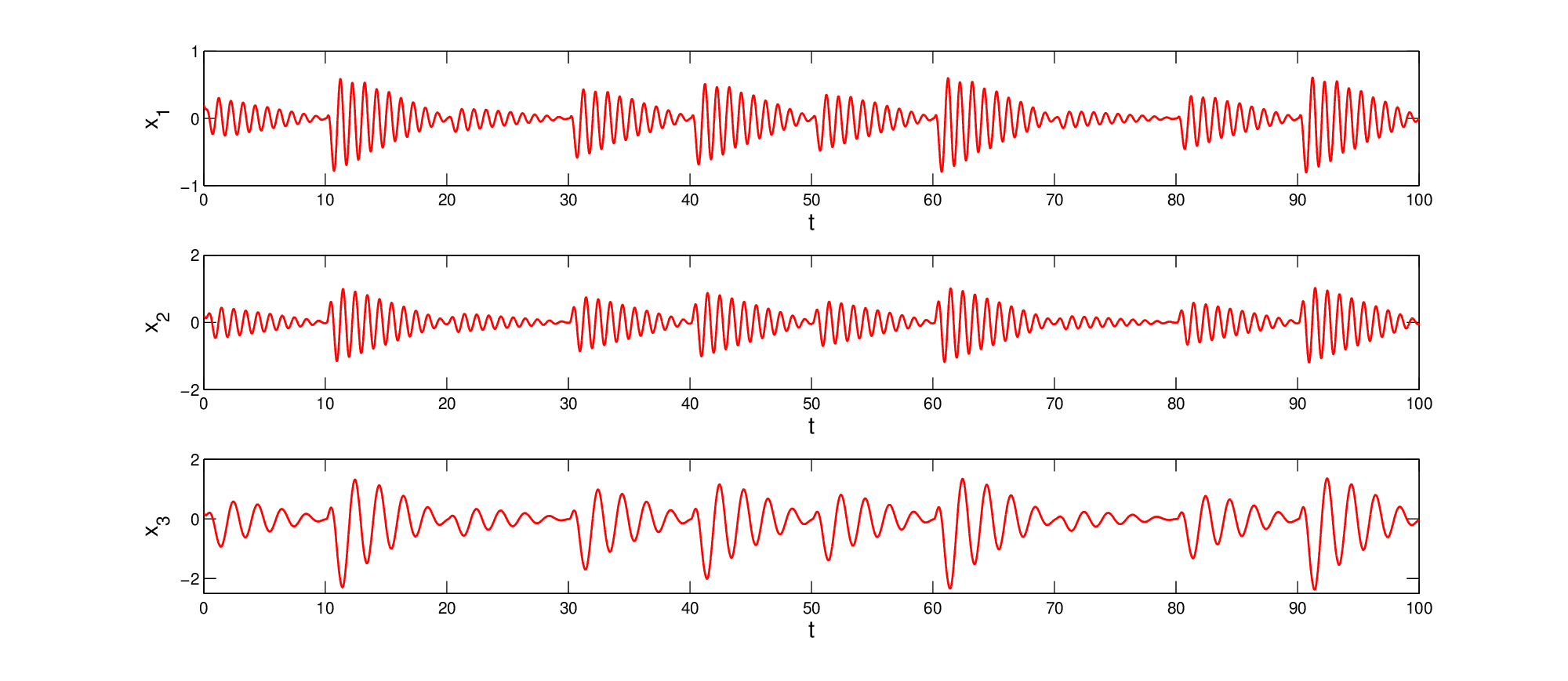}
	\caption{The coordinates of solution $x(t)$ of the neural network (\ref{ex1}), which approach the coordinates of unpredictable solution $z(t).$ The degree of periodicity is equal to 1/5.} \label{fig1}
\end{figure}

\begin{figure}[ht]
	\centering
	\includegraphics[width=9cm]{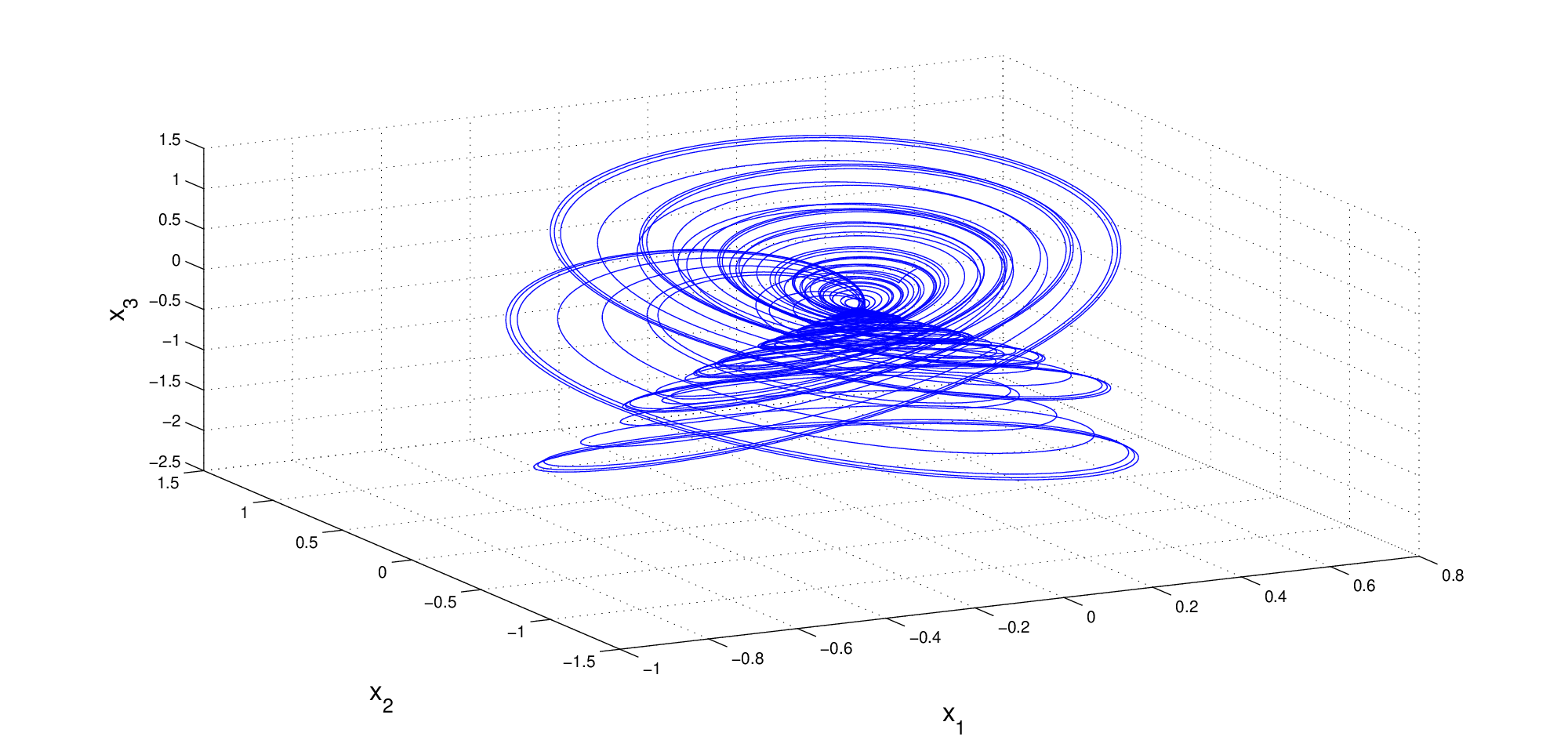}
	\caption{Trajectory of solution $x(t),$ which approximates the unpredictable solution $z(t).$ The degree of periodicity is equal to 1/5.} \label{fig2}
\end{figure}

\noindent \textbf{Example 2.} Let us take the neural networks (\ref{ex1}) with 
the same coefficients as in Example 1, with the only difference that the  strengths of connectivity and input data are compartmental periodic unpredictable:
$c_{11}(t)=0.1\sin(\pi t)\Theta(t),$ $c_{12}(t)=0.2\cos(4\pi t)\Theta(t),$ $c_{13}(t)=0.1\sin(2\pi t)\Theta(t),$ $c_{21}(t)=0.2\cos(\pi t)\Theta(t),$ $c_{22}(t)=0.05\sin(2\pi t)\Theta(t),$ $c_{23}(t)=0.1\cos(2\pi t)\Theta(t),$  $c_{31}(t)=0.05\sin(4\pi t)\Theta(t),$ $c_{32}(t)=0.1\cos(\pi t)\Theta(t),$ $c_{33}(t)=0.2\sin(4\pi t)\Theta(t),$ $v_1(t)=4\cos(2\pi t)\Theta(t),$ $v_2(t)=2\sin(2\pi t)\Theta(t),$ $v_3(t)=3\cos(\pi t)\Theta(t).$ The function $\Theta(t)$ is such that $\Theta(t)=\int_{-\infty}^{t}e^{-3(t-s)}\pi(s)ds,$ with  $\pi(t)=\lambda_i (t-i)$ if $t\in(10i, 10(i+1)].$ As we see, the periodic components are $2-$periodic, and the degree of periodicity, $\nabla,$ is equal to 1/5. The convergence sequence $t_p$ is a subsequence of numbers $10p,$ $p=0,1,2,\cdots,$ so it satisfies the kappa property. All conditions of Theorem \ref{theorem4} are satisfied. Figures \ref{fig1} and \ref{fig2} demonstrate the dynamics of (\ref{ex1}), with initial values $x_1(0)=x_2(t)=x_3(0)=0.2,$ and $\nabla=1/5.$

\begin{figure}[ht]
	\centering
	\includegraphics[width=9cm]{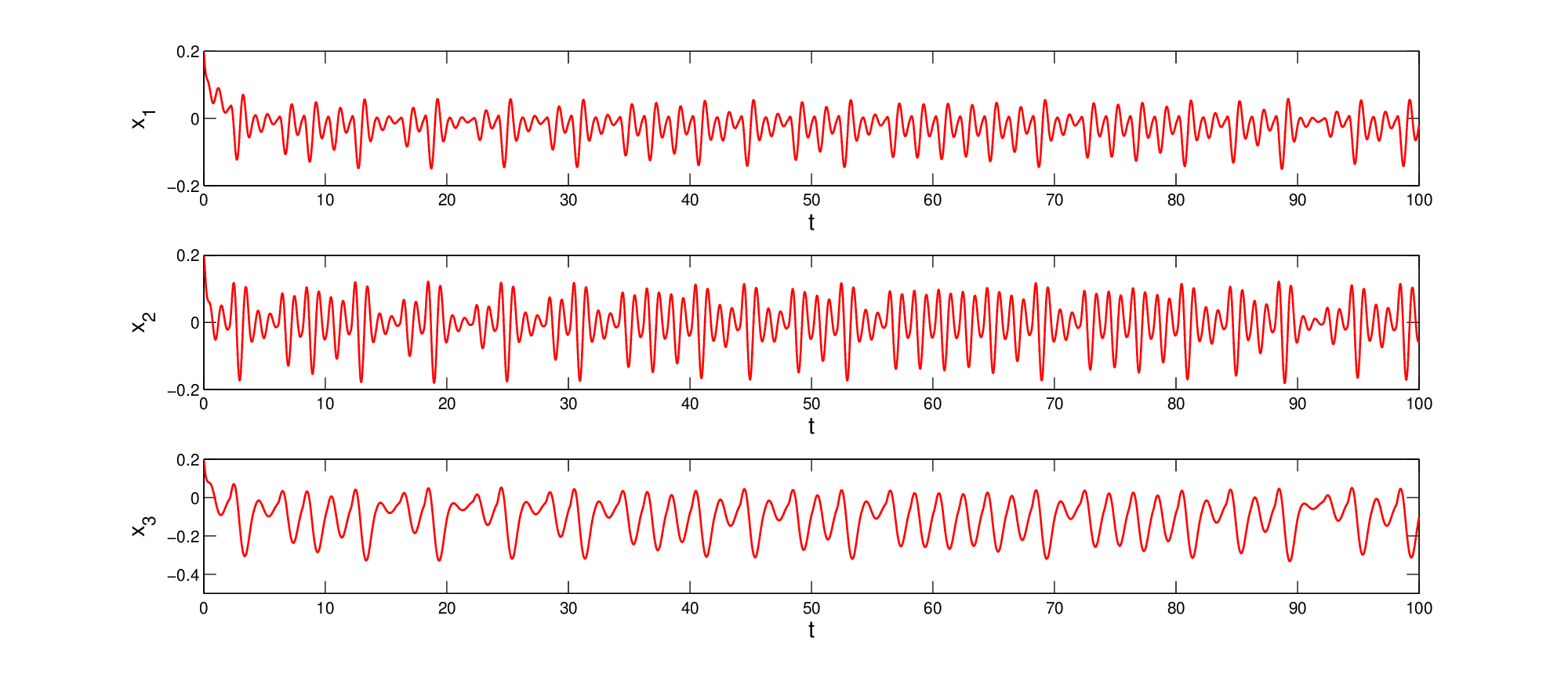}
	\caption{Dynamics of the coordinates of solution $x(t),$ which approach the coordinates of unpredictable solution $z(t).$ The degree of periodicity is equal to 1.} \label{fig3}
\end{figure}

\begin{figure}[ht]
	\centering
	\includegraphics[width=9cm]{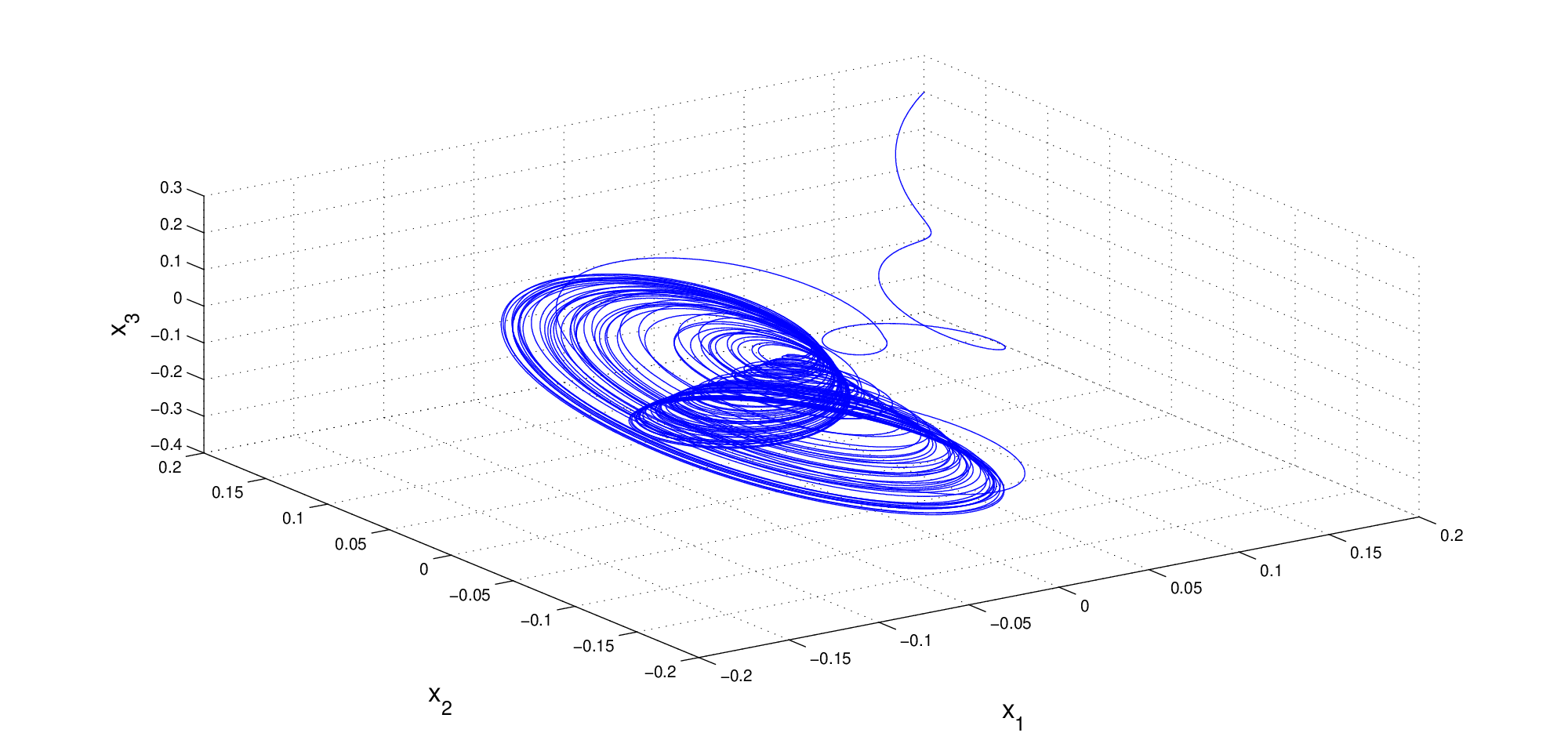}
	\caption{The irregular trajectory of solution $x(t)$ when the degree of periodicity is equal to 1.} \label{fig4}
\end{figure}

If $\Theta(t)=\int_{-\infty}^{t}e^{-3(t-s)}\pi(s)ds,$ where  $\pi(t)=\lambda_i (t-i)$ for $t\in(2i, 2(i+1)],$ then the degree of periodicity is equal to one. And we get irregular behaviour of neural network (\ref{ex1}), which is presented in Figures \ref{fig3} and \ref{fig4}.

\begin{figure}[ht]
	\centering
	\includegraphics[width=9cm]{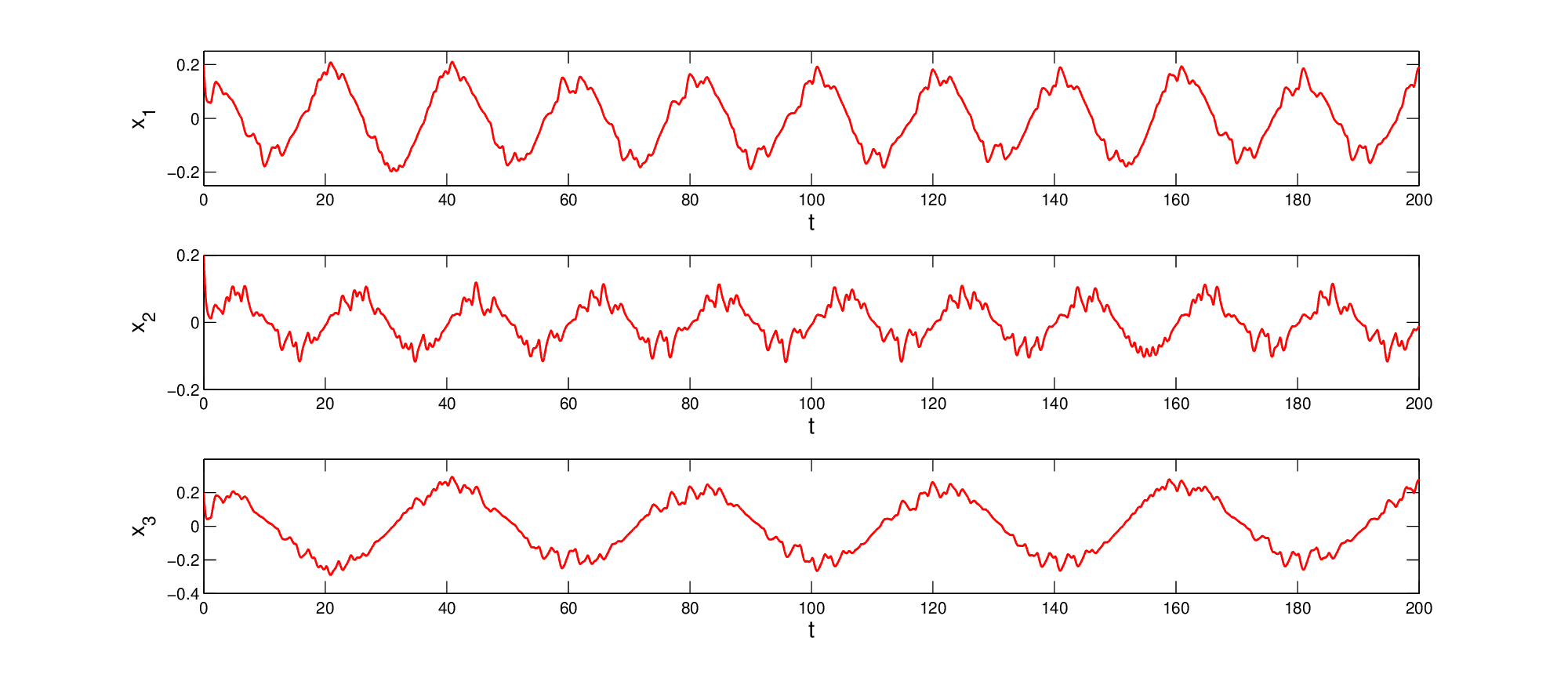}
	\caption{The time series of the coordinates $x_1(t), x_2(t)$ and $x_3(t)$ of the solution of system (\ref{ex1}), which exponentially converges to the coordinates of unpredictable solution. The degree of periodicity is equal to 40.} \label{fig5}
\end{figure}

\begin{figure}[ht]
	\centering
	\includegraphics[width=6cm]{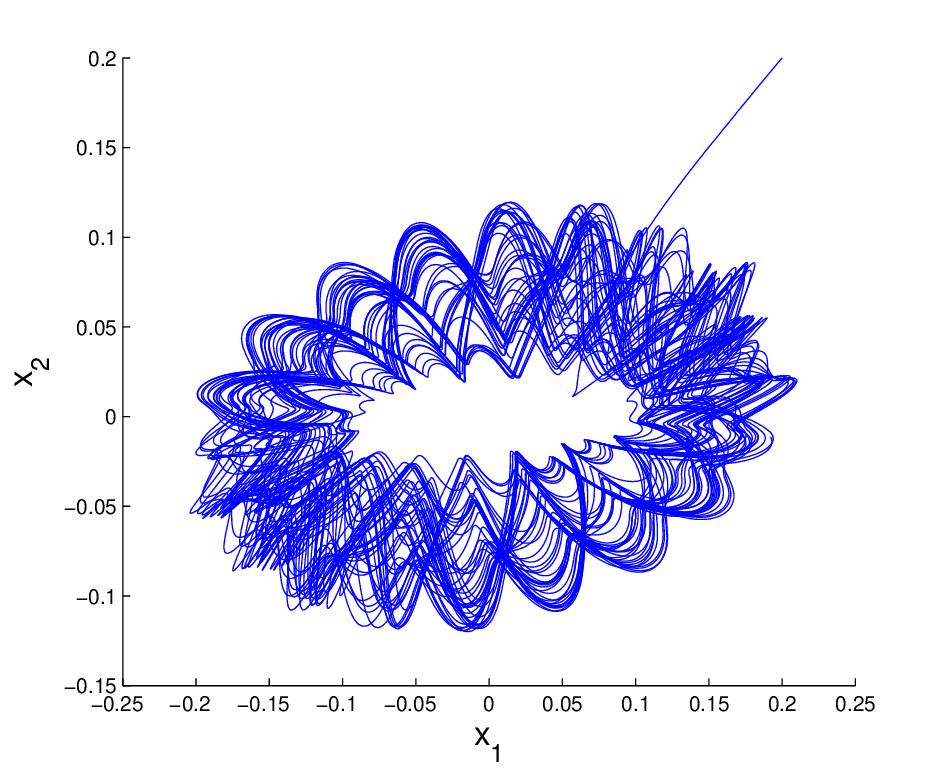}
	\caption{The projection of the trajectory of neural network (\ref{ex1}) on the $x_1-x_2$ plane for $t\in[0,1000].$ The degree of periodicity is equal to 40.} \label{fig6}
\end{figure}

\begin{figure}[ht]
	\centering
	\includegraphics[width=9cm]{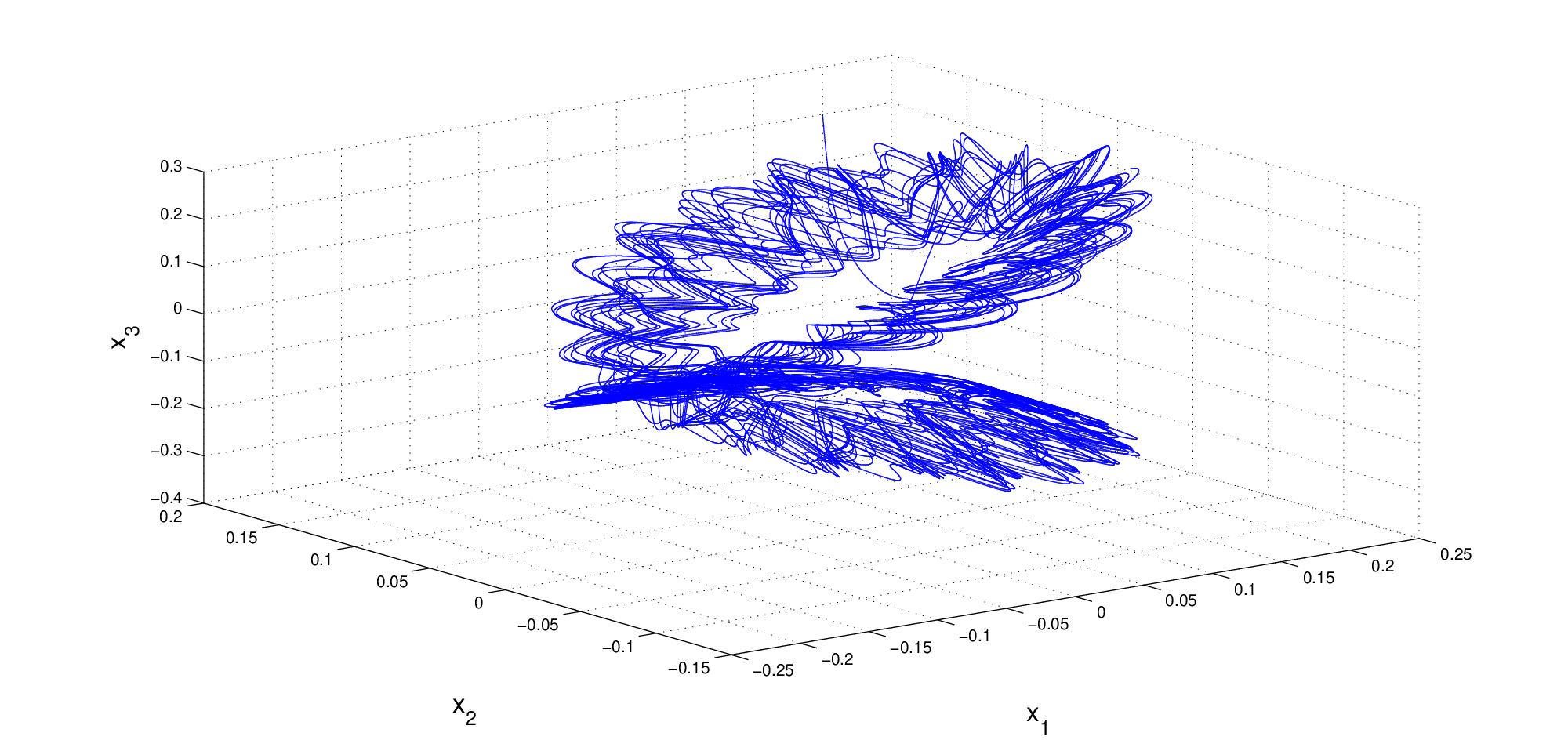}
	\caption{The trajectory of system (\ref{ex1}) for $t\in[0,1000]$ when the degree of periodicity is equal to 40.} \label{fig7}
\end{figure}

\noindent \textbf{Example 3.} In this part, we take compartmental periodic unpredictable functions with common $40-$periodic components such that
$c_{11}(t)=0.1\sin(0.1\pi t)\Theta(t),$ $c_{12}(t)=0.2\cos(0.4\pi t)\Theta(t),$ $c_{13}(t)=0.1\sin(0.2\pi t)\Theta(t),$ $c_{21}(t)=0.2\cos(0.1\pi t)\Theta(t),$ $c_{22}(t)=0.05\sin(0.2\pi t)\Theta(t),$ $c_{23}(t)=0.1\cos(0.2\pi t)\Theta(t),$ $c_{31}(t)=0.05\sin(0.4\pi t)\Theta(t),$ $c_{32}(t)=0.1\cos(0.1\pi t)\Theta(t),$ $c_{33}(t)=0.2\sin(0.4\pi t)\Theta(t),$ $v_1(t)=4\cos(0.1\pi t)\Theta(t),$ $v_2(t)=2\sin(0.1\pi t)\Theta(t),$ $v_3(t)=3\cos(0.05\pi t)\Theta(t).$ The function $\Theta(t)$ is determined on intervals $(i, i+1],$ and the degree of periodicity is equal to 40. Figure \ref{fig5} shows the time series of the coordinates $x_1(t),$ $x_2(t)$ and $x_3(t)$ of the solution $x(t)$ of (\ref{ex1}). The coordinates
$x_1-x_2$ and $x_1-x_2-x_3$ of the trajectory are demonstrated in Figures \ref{fig6} and \ref{fig7}, respectively.
Analyzing numerical examples for neural network with compartmental periodic unpredictable input data, it is possible to make interesting observations regarding the predominance of periodicity and unpredictability of outputs. Figure \ref{fig5} shows that if $\nabla>1$ then the graphs admit a clear periodic shape, which is enveloped by the unpredictability. In contrast, if $\nabla\leq 1,$ one can see in Figures \ref{fig1} and \ref{fig3} that the unpredictability prevails.

\section{Conclusions} 
In this paper we provide theoretical as well as numerical results for Poisson stable and unpredictable oscillations in CGNNs with variable unpredictable and compartmental periodic unpredictable strengths of connectivity between cells and inputs. Sufficient conditions were obtained to guarantee the existence of exponentially stable unpredictable and Poisson stable solutions. By numerical simulations, it is shown how a special technical characteristic, the degree of periodicity, effects to estimate contributions of periodic and unpredictable arguments to the behaviour of the neural network.  We  compared  Figures \ref{fig01},\ref{fig1},\ref{fig3} and \ref{fig5} with experimental data in papers \cite{ZhangLu,LiZh,Mohammad}, and it was   found that they are surprisingly similar. It means that the unpredictable functions can find applications in solutions of industrial problems. In addition, since the efficiency of neural networks strongly depends on the choice of input data \cite{Carpenter,Fukushima}, it
will be productive if the study of synchronization takes into account the compartmental periodic unpredictable functions. \cite{Zhang2021,Zhang2023,Das2000,Korn2003,Choi2019,Aihara1990}



\section*{Acknowledgements}

M. Akhmet and A. Zhamanshin have been supported by  2247-A National Leading Researchers Program of TUBITAK, Turkey, N 120C138. M. Tleubergenova has been supported by the  Committee of Science of the Ministry of Science and Higher Education of the Republic of Kazakhstan (grant AP14870835).

\section*{CRediT authorship contribution statement}

Marat Akhmet: Conceptualization, Methodology. Madina Tleubergenova: Investigation, Supervision, Writing–review and editing. Akylbek Zhamanshin: Software, Investigation, Writing - original draft. All authors have read and agreed to the published version of the manuscript.

\section*{Declaration of Competing Interest}

The authors declare that they have no known competing financial interests or personal relationships that could have appeared to influence the work reported in this paper.

\bibliographystyle{model1-num-names}

\bibliography{cas-refs}

\end{document}